\documentclass[preprint]{aastex63}
\usepackage{CJK}
\usepackage{soul}

\shortauthors{Yu et al.}
\accepted{May 2, 2021}
\submitjournal{ApJ}
\begin{document}

\title{How to identify exoplanet surfaces using atmospheric trace species in hydrogen-dominated atmospheres}

\correspondingauthor{Xinting Yu}
\email{xintingyu@ucsc.edu}

\author[0000-0002-7479-1437]{Xinting  Yu \begin{CJK*}{UTF8}{gbsn}(余馨婷)
\end{CJK*}}
\affiliation{Department of Earth and Planetary Sciences \\
University of California Santa Cruz \\
1156 High Street\\
Santa Cruz, California 95064, USA}

\author{Julianne I. Moses}
\affiliation{Space Science Institute, \\
Boulder, Colorado 80301, USA}

\author{Jonathan J. Fortney}
\affiliation{Department of Astronomy and Astrophysics \\
University of California Santa Cruz \\
1156 High Street\\
Santa Cruz, California 95064, USA}

\author{Xi Zhang}
\affiliation{Department of Earth and Planetary Sciences \\
University of California Santa Cruz \\
1156 High Street\\
Santa Cruz, California 95064, USA}

\begin{abstract}
Sub-Neptunes (R$_p\sim$1.25--4 R$_{\mathrm{Earth}}$) remain the most commonly detected exoplanets to date. However, it remains difficult for observations to tell whether these intermediate-sized exoplanets have surfaces and where their surfaces are located. Here we propose that the abundances of trace species in the visible atmospheres of these sub-Neptunes can be used as proxies for determining the existence of surfaces and approximate surface conditions. As an example, we used a state-of-the-art photochemical model to simulate the atmospheric evolution of K2-18b and investigate its final steady-state composition with surfaces located at different pressures levels (P$_{\mathrm{surf}}$). We find the surface location has a significant impact on the atmospheric abundances of trace species, making them deviate significantly from their thermochemical equilibrium and ``no-surface" conditions. This result arises primarily because the pressure-temperature conditions at the surface determine whether photochemically-produced species can be recycled back to their favored thermochemical-equilibrium forms and transported back to the upper atmosphere. For an assumed H$_2$-rich atmosphere for K2-18b, we identify seven chemical species that are most sensitive to the existence of surfaces: ammonia (NH$_3$), methane (CH$_4$), hydrogen cyanide (HCN), acetylene (C$_2$H$_2$), ethane (C$_2$H$_6$), carbon monoxide (CO), and carbon dioxide (CO$_2$). The ratio between the observed and the no-surface abundances of these species, can help distinguish the existence of a shallow surface (P$_{\mathrm{surf}}<$ 10 bar), an intermediate surface (10 bar $<\mathrm{P}_{\mathrm{surf}}<$ 100 bar), and a deep surface (P$_{\mathrm{surf}}>$ 100 bar). This framework can be applied together with future observations to other sub-Neptunes of interest.

\end{abstract}


\section{Introduction} \label{sec:intro}
The Kepler mission has detected a wealth of exoplanets that do not resemble any planetary bodies in the Solar System, with sizes in between Earth and Neptune, 1.0 R$_{\mathrm{Earth}}<\mathrm{R}_p<$ 3.9 R$_{\mathrm{Earth}}$ \citep{2013ApJ...766...81F,2014PNAS..11112647B}. For these intermediate-sized exoplanets, it is unclear if they are closer to the terrestrial planets, so-called ``super-Earths", where the gas-solid or gas liquid interface is located at a shallow pressure level (e.g., $<$ 100 bar) or if they resemble scaled-down versions of the ice giants, so-called ``mini-Neptunes", with a gas-solid/liquid interface (if there is any) located deeply at high pressure levels (e.g., $>$ 1 kbar). 

Population studies have found a gap in exoplanet occurrence between 1.5--2.0 R$_{\mathrm{Earth}}$ \citep{2017AJ....154..109F,2018AJ....156..264F,2018MNRAS.479.4786V}, which could be explained by mass-loss processes such as photoevaporation \citep{2013ApJ...775..105O,2013ApJ...776....2L} or core-powered mass loss \citep{2018MNRAS.476..759G}. These mass-loss theories suggest that highly irradiated intermediate-sized planets are expected to have lost any primordial H/He atmospheric envelope, becoming primarily airless rocky cores that have smaller observed radii, while the less irradiated planets could retain their primordial hydrogen (H$_2$)-rich atmospheres, such that their observed radii are inflated to values greater than 2.0 R$_{\mathrm{Earth}}$. The population of exoplanets with radii larger than 2 R$_{\mathrm{Earth}}$ can also be explained by the volatile-rich interior structures of these exoplanets, such that variations in their intrinsic water content can reproduce the observed radius peak at around 2.5 R$_{\mathrm{Earth}}$ \citep{2019PNAS..116.9723Z, 2020ApJ...896L..22M,2021JGRE..12606639B}. For planets with radii greater than 3.0 R$_{\mathrm{Earth}}$, \citet{2019ApJ...887L..33K} proposed that the base-of-atmosphere pressures are likely large enough that their atmospheres readily dissolve into the magma ocean, which inhibits further growth to planets of larger sizes.

In this study, we are interested in investigating exoplanets with radii in the $\sim$1.2-3 R$_{\mathrm{Earth}}$ range that are within or at the high-radius side of the radius gap valley, such that they may not have lost all their original atmospheric envelope \citep[e.g.,][]{2020A&A...638A..52M}. The exoplanets of interest to our study are those with small atmospheric mass fractions, f$_{\mathrm{atm}}$. Following Equation 1 in \citet{2019ApJ...887L..33K}, for a 1-bar surface, f$_{\mathrm{atm}}$ is $\sim10^{-7}$ to $\sim10^{-5}$, and for a 100-bar surface, f$_{\mathrm{atm}}$ is around $10^{-5}$ to $10^{-3}$. For reference, Earth's atmospheric fraction is $\sim8\times10^{-7}$. Through a combination of our theoretical predictions and future observations, we hope to be able to distinguish whether these intermediate planets are more akin to rocky terrestrial bodies with shallow surfaces (or so called ``super-Earths"), or highly irradiated gas giants with deep and hot atmospheres (or so-called ``mini-Neptunes"). The location of an underlying solid surface has significant impacts on the abundances of trace species in the atmospheres of Solar System bodies, primarily because the pressure-temperature conditions at this lower boundary determine whether the photochemically-formed species can be recycled back to the atmosphere to replenish the ``parent" atmospheric species \citep[e.g.,][]{1969JAtS...26..906S,1973JAtS...30.1205S}. Thus, we propose that atmospheric abundances of trace species may be used as proxies to probe for the existence of surfaces on sub-Neptunes.

The existence of a surface may lead to particular atmospheric composition dichotomies between the giant planets and terrestrial planets in the Solar System. For example, ammonia (NH$_3$) is an important minor constituent in the atmospheres of the giant planets in the Solar System \citep{2009book....156..264F}. In the upper troposphere of Jupiter, NH$_3$ is irreversibly destroyed to form nitrogen compounds such as hydrazine (N$_2$H$_4$, which condenses) and nitrogen \citep[N$_2$,][]{1973JAtS...30.1205S}. However, ammonia remains abundant in the Jovian atmosphere, as the N$_2$H$_4$ and N$_2$ can be transported into the deep, hot part of the atmosphere and be converted back to NH$_3$ through thermochemistry \citep[e.g.,][]{1973JAtS...30.1205S}.

In contrast, the NH$_3$ abundance is very low in the atmosphere of Saturn's moon Titan \citep[volume mixing ratio, VMR$<10^{-10}$ in the stratosphere,][]{2010FaDi..147...65N,2013P&SS...75..136T}\footnote{Ammonia has a volume mixing ratio of $7\times10^{-6}$ in Titan's ionosphere \citep{2007Icar..191..722V, 2009P&SS...57.1558V}, due to ion chemistry in Titan's upper atmosphere \citep{2010FaDi..147...31Y}.}, even though ammonia is likely one of the primordial ingredients that Titan originally accreted \citep{2014ApJ...788L..24M}. Ammonia is irreversibly destroyed by photochemistry in Titan's upper atmosphere and is likely the ultimate source of Titan's present nitrogen \citep[e.g.,][]{2010tfch.book..177A}. The lack of ammonia on Titan is thus likely the result of the existence of a shallow surface on Titan, which prevents thermochemical recycling of the N$_2$ back into NH$_3$, such as what would have occurred in the hot, deep part of the atmosphere of a planet with a thicker atmosphere.

Similar to NH$_3$, it is not surprising to find methane (CH$_4$) in the atmospheres of the giant planets \citep{2009book....156..264F}, but quite unexpected for Titan. With no recycling from thermochemistry, CH$_4$ in Titan's atmosphere should be irreversibly destroyed by photochemistry to form more complex hydrocarbons in $\sim$ 10 Myrs, as determined by \citep{1984ApJS...55..465Y}. While the source of the present-day methane on Titan is still a mystery \citep[e.g.,][]{1987Icar...70...61L, 2006P&SS...54.1177A, 2015Icar..250..570G, 2018NatGe..11..306H}, the Huygens Probe measurements of primordial noble gases indicate that Titan's current atmosphere is linked to its interior rather than accreted upon formation \citep{2005Natur.438..779N, 2010JGRE..11512006N}, and the current amount of methane is likely a result of outgassing from Titan's interior \citep{2006Natur.440...61T}.

Inspired by these two types of bodies in the Solar System, we propose that the abundances of atmospheric trace species such as ammonia, methane, or others could be the proxies for surface identification on exoplanets. The upcoming exoplanet spectroscopy missions such as the James Webb Telescope (JWST) and the Atmospheric Remote-sensing Infrared Exoplanet Large-survey (ARIEL) would be excellent in characterizing atmospheric composition of exoplanets with near- to mid-infrared spectra, but transit observations may not be useful in characterizing surfaces, and infrared emission observations may not be able to probe down to the surface if the atmosphere is too thick. Thus, identifying potential observable atmospheric species that could be used to point to the existence of exoplanet surfaces would be an intriguing science angle for the two missions.

The paper is structured as follows. We use temperate sub-Neptune K2-18b as our model planet and investigate its atmospheric chemical evolution with and without surfaces. Previous works on understanding the properties of K2-18b are summarized in Section 2.1. The thermochemical and photochemical kinetics model is described in Section 2.2. The final steady-state compositions for K2-18b and a hotter variant of K2-18b, both with and without surfaces, are shown in Section 3.1. The sensitivity of species to different surface levels are summarized and discussed in Section 3.2. In Section 3.3, we examine the sensitivity of our results to a few planetary parameters. We discuss the potential applications of our results for future observations in Section 4.

\section{Methods} \label{sec:methods}
\subsection{K2-18b}
While our modeling is designed to represent generic sub-Neptunes, we adopt the properties of K2-18b for our models. K2-18b is an intermediate-sized sub-Neptune \citep[radius 2.71 R$_{\mathrm{Earth}}$,][]{2019A&A...621A..49C} in the habitable zone of a moderately active early M-dwarf star \citep[M2.5 $\pm$ 0.5,][]{2017A&A...608A..35C}, with equilibrium temperature of 255 $\pm$ 4 K (assuming an albedo of 0.3). The atmosphere of K2-18b has been characterized by the Kepler Space Telescope (K2) in the 0.4-0.9 $\mu$m range, the Hubble Space Telescope (HST) Wide Field Camera 3 (WFC3) in the 1.1-1.7 $\mu$m range, and the Spitzer Telescope at 3.6 and 4.5 $\mu$m \citep{2017ApJ...834..187B, 2019NatAs...3.1086T, 2019ApJ...887L..14B}. In contrast to the flat spectra observed for most sub-Neptunes \citep{2014Natur.505...69K, 2014Natur.505...66K, 2014ApJ...794..155K, 2020AJ....159...57L, 2020AJ....159..239G, 2020AJ....160..201C}, the observations of K2-18b reveal a significant absorption feature at 1.4 $\mu$m possibly due to water (H$_2$O) or methane, assuming a H$_2$/He-dominated atmosphere \citep{2019NatAs...3.1086T, 2020arXiv201110424B, 2020ApJ...898...44S}, albeit with some evidence for water ice/liquid clouds \citep[e.g.,][]{2019ApJ...887L..14B, 2020ApJ...904..154P, 2021A&A...646A..15B,2021a&a201111553C}. 

Even though a high-mean-molecular-weight atmosphere (H$_2$O-dominated) with some N$_2$ and/or H$_2$/He cannot be ruled out given the current HST/WFC3 data \citep{2019NatAs...3.1086T}, photoevaporation models and Lyman alpha observations \citep[e.g.,][]{2020A&A...638A..52M, 2020A&A...634L...4D} suggest that K2-18b could have retained a large percentage of any primordial H/He atmospheric envelope due to the combination of its moderate size and relatively low stellar irradiation levels. The mass and radius of K2-18b \citep{2019A&A...621A..49C} are also fully consistent with a H$_2$-rich, Neptune-like atmosphere. On the other hand, atmosphere and interior models \citep{2020ApJ...891L...7M, 2020ApJ...904..154P} suggests that K2-18b could have a cool, habitable liquid water surface (T $<$ 400 K) with pressure as low as 1 bar. However, identifying the existence of a surface and retrieving the surface pressure of K2-18b would likely be difficult with solely JWST/ARIEL transit observations \citep{2020arXiv200301486C}.

\subsection{Thermo/photochemical model}
Here we use an established one-dimensional (1D) thermochemical and photochemical kinetics and transport model, based on the Caltech/JPL KINETICS code \citep{1980ApJ...242L.125A, 2011ApJ...737...15M, 2013ApJ...777...34M, 2016ApJ...829...66M, 2014RSPTA.37230073M} to simulate the atmospheric evolution of K2-18b. This model bridges the photochemistry-dominated upper atmosphere and the thermochemistry-dominated deep atmosphere seamlessly and considers potential disequilibrium chemical processes such as transport-induced quenching and photochemistry \citep{2011ApJ...737...15M}, allowing us to investigate the effect of different surface pressure (P$_{\mathrm{surf}}$) levels on the chemical evolution of major observable species containing C, H, O, N in the atmosphere of K2-18b. We run the model with surfaces located at four different pressure levels (1-bar, 10-bar, 100-bar, and deep/no surface). The model runs until a steady state is reached for a convergence criterion of one part in a thousand.

The required inputs to the model include: 1) the assumed bulk elemental composition of the atmosphere, 2) the atmospheric pressure-temperature profile (P-T profile), 3) the initial atmospheric composition along this P-T profile, which is assumed to be in thermochemical equilibrium throughout, 4) the stellar spectrum, 5) the vertical profile of eddy diffusion coefficients K$_{zz}$, 6) constituent boundary conditions, and 7) chemical inputs such as species' ultraviolet cross sections and reaction rate coefficients.

We assume a hydrogen-dominated but metal-rich atmosphere for the sub-Neptune K2-18b, with a metallicity of 100 times solar \citep[where the ``solar composition" is defined from the protosolar abundances of][]{2010ASSP...16..379L}, which is within the metallicity range of the best-fit HST data \citep{2020arXiv201110424B, 2021A&A...646A..15B, 2021a&a201111553C} and is plausible from formation models \citep{2013ApJ...775...80F}. The P-T profiles in our model are generated with a well-established 1D radiative-convective model \citep{1999Icar..138..268M, 2005ApJ...627L..69F, 2008ApJ...678.1419F, 2012ApJ...756..172M}. The model accounts for both incident radiation from the parent star and thermal radiation from the planet's atmosphere and interior. The radiative transfer scheme is described in \citep{1989JGR....9416287T} and \citep{1989Icar...80...23M}, with further model details and methods described in \citep{1989Icar...80...23M}.

\begin{figure}[h]
\centering
\includegraphics[width=\textwidth]{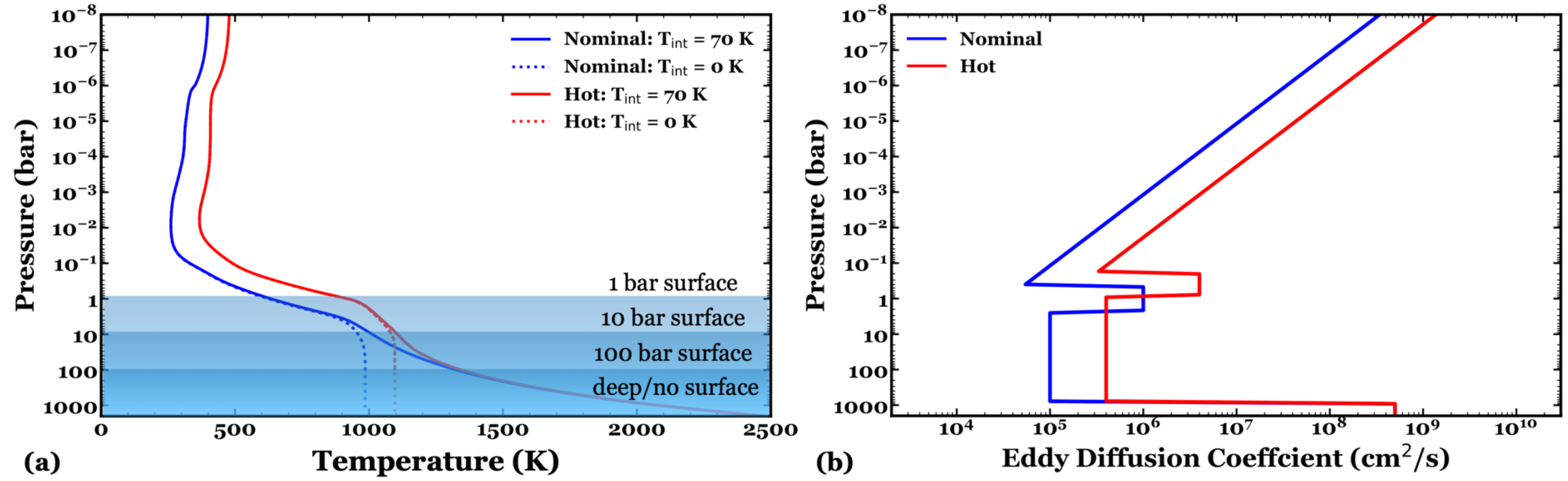}
\caption{(a) The pressure-temperature profiles used in our model, based on the 1D radiative-convective model of \citet{2005ApJ...627L..69F, 2008ApJ...678.1419F}. The blue lines show the nominal profile for K2-18b and the red lines are profiles for a hypothetical hotter sub-Neptune with four times the nominal stellar flux of K2-18b. The solid lines are P-T profiles with intrinsic temperatures T$_{\mathrm{int}}$ of 70 K and the dotted lines are for the T$_{\mathrm{int}}$ = 0 K cases. (b) The nominal eddy diffusion coefficient profiles used in our model for the nominal K2-18b and the hotter variant. Note that a detached convective zone is found around 1 bar in both cases.}
\label{fig:TP}
\end{figure}

Here we assume a moderate 100 times solar metallicity and use two end-member intrinsic fluxes from the planetary interior ($F_{\mathrm{int}} = \sigma T_{\mathrm{int}}^4$), parameterized by the intrinsic temperature T$_{\mathrm{int}}$, of respectively 0 K and 70 K to calculate the P-T profile. The intrinsic flux depends on the size, age, and composition of the planet. T$_{\mathrm{int}}$ values for Neptune and Saturn are 53 K and 78 K, respectively \citep{1991JGR....9618921P}. \citet{2013ApJ...776....2L} suggest that T$_{\mathrm{int}}$ values of $\sim$30 K could be appropriate for a planet with K2-18b's properties, so the T$_{\mathrm{int}}$ = 0 and 70 K end-member cases should bracket the likely range of possibilities. Because the model-generated P-T profiles only go to 1 $\mu$bar at high altitudes, and important photochemistry occurs at altitudes above that level, we extended the profiles to lower pressures by assuming an arbitrary power-law expression that extends to a nearly isothermal profile at pressures $<$ 1 $\mu$bar. The thermo/photochemical kinetics model also requires that the deep atmosphere extends to temperatures of $\sim$2600 K to accurately capture the quench points for all species so they can reach thermochemical equilibrium at depth. For the T$_{\mathrm{int}}$ = 70 K case, we extended the model-derived temperature profile to deeper levels assuming an adiabat. The T$_{\mathrm{int}}$ = 0 K profile reaches a deep isothermal temperature that is cooler than 2600 K and so was not extended to greater depths, thus, some species (e.g., nitrogen species N$_2$-NH$_3$) cannot achieve full thermochemical equilibrium at any point in the atmosphere for this T$_{\mathrm{int}}$ = 0 K case.  To examine a phase space for warmer sub-Neptune planets, we also generated a warmer P-T profile for a hypothetical hotter sub-Neptune by halving the current orbital distance of K2-18b, resulting in a factor of four increase in the solar insolation the planet receives. The final P-T profiles, shown in Figure \ref{fig:TP}(a), extend to pressures of 10-11 bar. For the model runs with different surface pressure levels, we simply cut the P-T profiles at each corresponding surface level (1-bar, 10-bar, or 100-bar).

Once the P-T profiles for the different cases are established, we generate an equilibrium initial composition for the atmosphere using the NASA CEA thermochemical-equilibrium model of \citep{1984cpcc.book.....G}. Since the ultraviolet stellar spectrum of K2-18 (M2.5) has not been characterized yet, here we use the stellar spectrum of a moderately-active M-dwarf of similar spectral type, GJ 436 (M2.5) from \citep{2013ApJ...777...34M} \citep[see also the updated GJ 436 spectrum from the MUSCLES database,][]{2016ApJ...820...89F}. 

The eddy diffusion coefficient profiles are shown in Figure \ref{fig:TP}(b) for the nominal K2-18b and the hotter sub-Neptune variant. The K$_{zz}$ in the deep, convective portion of the atmosphere is estimated using the free-convection and mixing-length theories \citep[e.g.,][]{1976jupiter...26..906S}. For the upper atmosphere, K$_{zz}$ is assumed to vary inversely with the square root of atmospheric pressure, with a scaling that depends on the atmospheric scale height H(z) and the equilibrium temperature T$_{\mathrm{eq}}$, using the empirical formula described in \citet{2020ptrsa200611367M}. The temperature profiles predicted from the radiative-transfer models exhibit a detached convective region lying between two radiative regions. Based on the inferred K$_{zz}$ values for Earth and the Solar-System planets \citep[e.g.,][]{1999ppa..conf.....Y}, we assume a K$_{zz}$ of $10^6\ cm^2\ s^{-1}$ for the detached upper-tropospheric convective region in the cooler nominal model and a K$_{zz}$ of $10^5\ cm^2\ s^{-1}$ for the deeper radiative region in the upper troposphere. For the hotter planetary model, we assume K$_{zz}$ $=4\times10^6\ cm^2\ s^{-1}$ for the detached upper-tropospheric convective region and a K$_{zz}$ $=4\times10^5\ cm^2\ s^{-1}$ for the radiative region below it. However, given the large expected uncertainties in the K$_{zz}$ profile, we also test the sensitivity of our results to two additional K$_{zz}$ profiles for the nominal case, by multiplying and dividing the nominal K$_{zz}$ profile by a factor of three, hereinafter called the large and the small K$_{zz}$ cases. 

We assume zero flux for all species at both the top and bottom boundaries, which essentially assumes no sources or sinks at the surface (if there is one in the model) and no escape or influx at the top of the atmosphere. Although some atomic H and potentially other gases will be escaping under conditions relevant to K2-18b, \citet{2020A&A...634L...4D} demonstrate that the expected loss of H over the age of the system amounts to $<$ 1\% of the mass of the planet. Although volcanism, interior outgassing, and other geological or biological sources can provide potential non-zero flux terms to introduce new material to the atmosphere, and although chemical weathering, dissolution in oceans, sequestering of condensates and other surface-atmospheric interactions could provide potential sinks for atmospheric gases, the magnitude of such processes is unknown for exoplanets. The zero-flux boundary conditions therefore provide a good first-order prediction of the influence of the surface on the atmospheric composition in the absence of geological/biological sources and sinks.  We return to this point in the Discussion.

The fully reversed chemical reaction list of \citet{2013ApJ...777...34M} is adopted for this work. The model considers 92 neutral species that interact with each other through $\sim$1650 kinetic reactions, including hydrocarbons with molecular weight up to benzene (C$_6$H$_6$), nitriles and other nitrogen-bearing species with two or fewer N atoms, and oxygen species with three or fewer oxygen atoms. The non-photolysis reactions are fully reversed, allowing thermochemical equilibrium to be reproduced kinetically in the deep atmosphere when temperatures are large enough. However, when vertical transport is faster than the kinetic reactions can maintain that thermochemical equilibrium, certain constituents can have their mixing ratios ``quenched" \citep[e.g.,][]{1977Sci...198.1031P, 2002Icar..155..393L, 2011ApJ...737...15M,2012A&A...546A..43V} at values not representative of the local thermochemical equilibrium abundances. Photochemistry can further alter the abundances. Condensation is not included in this version of the model, which could potentially impact both the steady-state gas abundance profiles (e.g., H$_2$O would be overestimated at any altitudes where it condenses) and the time-variable atmospheric evolution (e.g., the oxygen tied up in the condensed H$_2$O could slow the speed of conversion of that oxygen into other gas-phase species, such as CO). Further details of the chemical model can be found in \citet{2005JGRE..110.8001M, 2011ApJ...737...15M, 2013ApJ...777...34M, 2016ApJ...829...66M}.

\section{Results}
\subsection{Photochemical model results}

\begin{figure}[h]
\centering
\includegraphics[width=\textwidth]{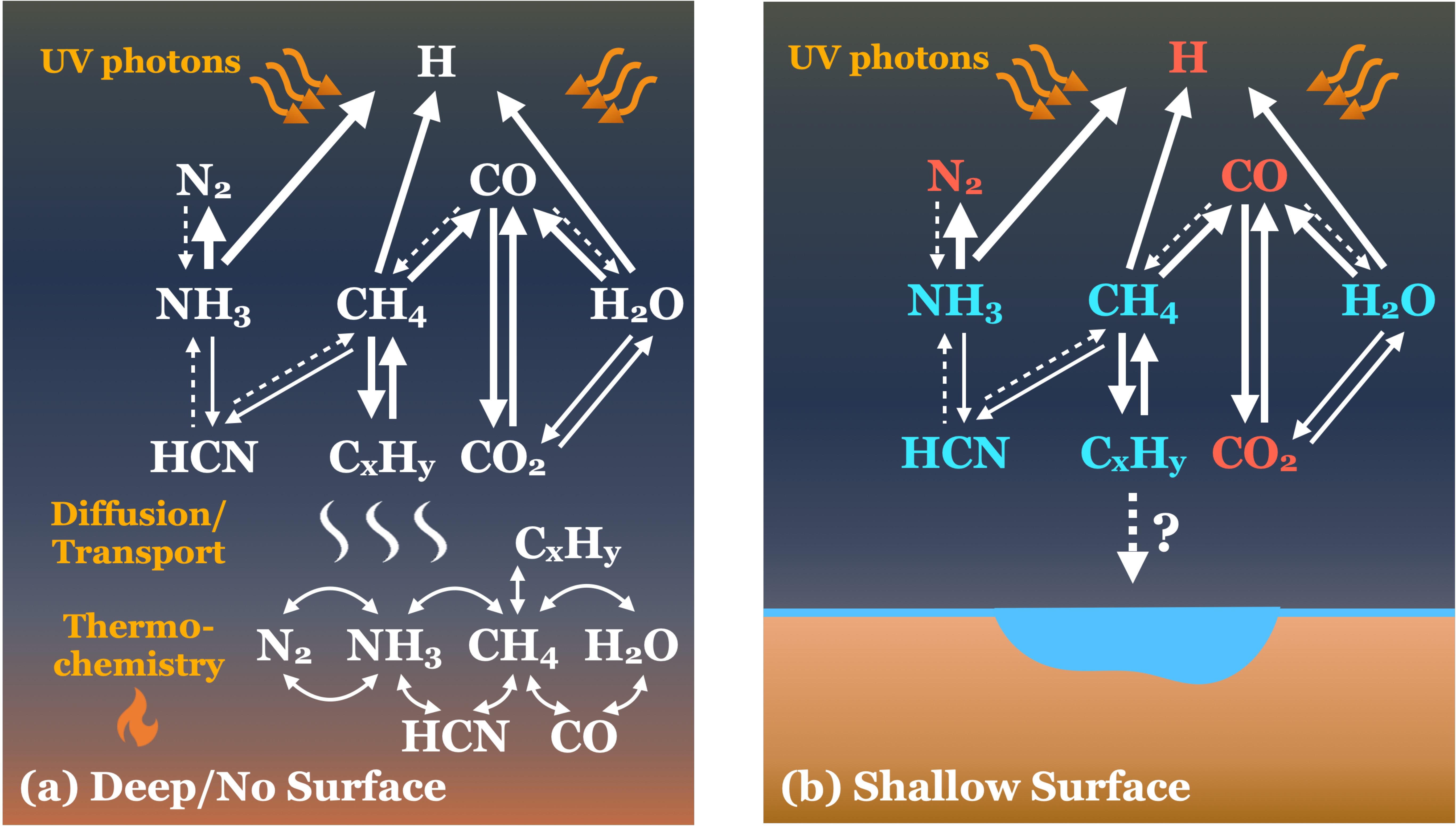}
\caption{Schematic diagram describing the main chemical pathways for the net production and loss of important observable species in the deep-surface or no-surface case versus the shallow-surface case for K2-18b. Arrows of different thicknesses indicate levels of importance for each chemical pathway, from high to low importance: thick solid arrows, thin solid arrows, dashed arrows. The upper atmosphere at pressures less than a few bars is dominated by photochemistry and vertical diffusion/transport, the troposphere from a few bars to the deep quench points (pressure depending on species involved) is dominated by vertical diffusion and transport-induced quenching, and the deep atmosphere below the quench points (provided that the atmosphere is hot enough) is dominated by thermochemical equilibrium. For the shallow, cool surface case, temperatures never get high enough for thermochemistry to be effective. The dashed arrow from C$_x$H$_y$ to the surface indicates potential condensation/sedimentation of refractory hydrocarbons on the surface of the exoplanet, which is a process that could happen if the surface temperatures are cold enough, but is not currently occurring in our model because none of the hydrocarbons considered are refractory enough to condense for the atmospheric and surface conditions of the planets considered here. The colored species indicate abundance fluctuations compared to the deep/no surface case, red means an increase in abundance and blue means a decrease in abundance.}
\label{fig:scheme}
\end{figure}

In Figure \ref{fig:scheme}, we summarize the main chemical pathways for the net production and loss of key species in the deep-surface/no-surface case and a shallow-surface case for K2-18b. For the no-surface case (Figure \ref{fig:scheme}a), the atmosphere can be divided into three sections that are dominated by different chemical processes. The upper region of atmosphere (pressures less than a few bars, depending on the thermal structure) is dominated by photochemistry and vertical diffusion, and the deepest, hottest part of atmosphere is dominated by thermochemical equilibrium (the pressures at which equilibrium dominates depend on the thermal structure of the atmosphere). The composition in the rest of the atmosphere is mainly controlled by species diffusion and transport; in this intermediate region, a process called transport-induced quenching, which occurs when transport timescales are shorter than the chemical kinetic reaction timescales, can act to prevent the constituents from achieving thermochemical equilibrium \citep[e.g.,][]{2002Icar..155..393L, 2011ApJ...738...72V}. The P-T point at which the transport timescale is equal to the chemical kinetic conversion timescale of a particular constituent is called that species' quench point. At altitudes below the quench point, the mixing ratio of a species will be governed by thermochemical equilibrium. At altitudes above the quench point, the mixing ratio of a species will remain constant with altitude at thermochemical-equilibrium value it reached at the quench point, unless other disequilibrium chemical processes such as photochemistry act upon it.

For the no-surface case, all three chemical processes conspire to determine the abundances profiles of species. In this scenario, species with weak bonds such as NH$_3$, CH$_4$, and H$_2$O are readily broken apart (photolyzed) by ultraviolet photons. The photolysis products undergo further chemical reactions that act to either reform the original ``parent" species or to form new photochemical products, such as N$_2$, hydrocarbons, nitriles, carbon monoxide (CO), and carbon dioxide (CO$_2$). When they are transported to the deep, hot part of the atmosphere, the photochemically-formed species will be converted back in part to NH$_3$, CH$_4$, and H$_2$O and transported back to the upper atmosphere. If a surface is located below the quench point of a species, thermochemistry can act to fully recycle this species back to the upper atmosphere as if there were no surface. However, if there exists a cold and shallow surface vertically above the point at which thermochemical equilibrium would have been achieved kinetically, then the chemical reactions do not work effectively to recycle the photochemical products back into the parent molecules. Thus, over time, the shallow-surface case would evolve to have decreased abundances of photochemically-fragile species (NH$_3$, CH$_4$, and H$_2$O, marked in blue in Figure \ref{fig:scheme}b) and increased abundances of more photochemically stable species (N$_2$, CO, and CO$_2$, marked in red in Figure \ref{fig:scheme}b) compared to the no-surface case.

\begin{figure}[h]
\centering
\includegraphics[width=\textwidth]{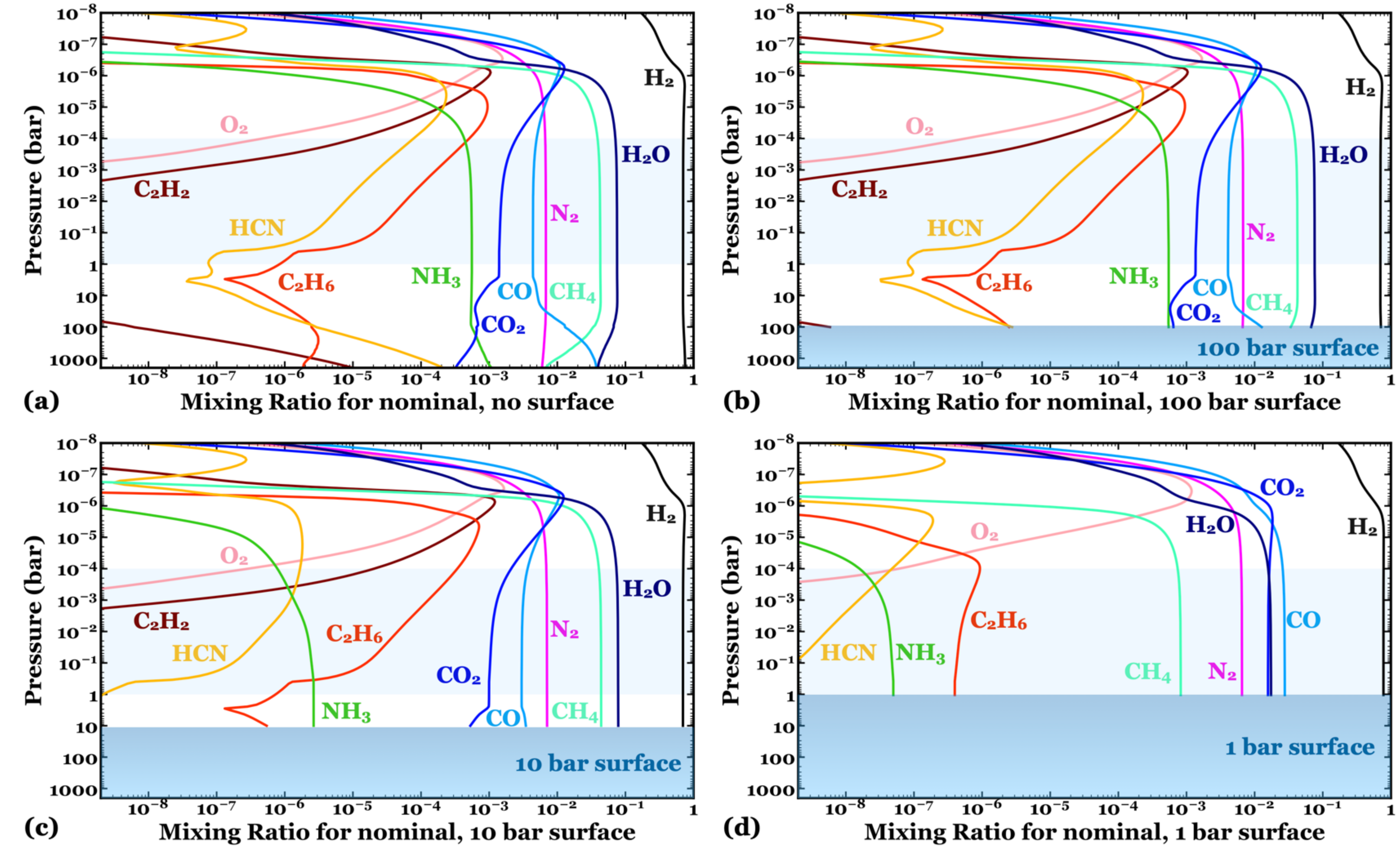}
\caption{Volume mixing ratio profiles for the main chemical species for the nominal K2-18b (blue P-T profile in Figure \ref{fig:TP}a) with T$_{\mathrm{int}}$ = 70 K, (a) without a surface (no-surface case) and with surface levels located at (b) 100-bar, (c) 10-bar, and (d) 1-bar levels. The shaded blue region indicates the observable part of the atmosphere (10-4 - 1 bar).}
\label{fig:cold}
\end{figure}

\begin{figure}[h]
\centering
\includegraphics[width=\textwidth]{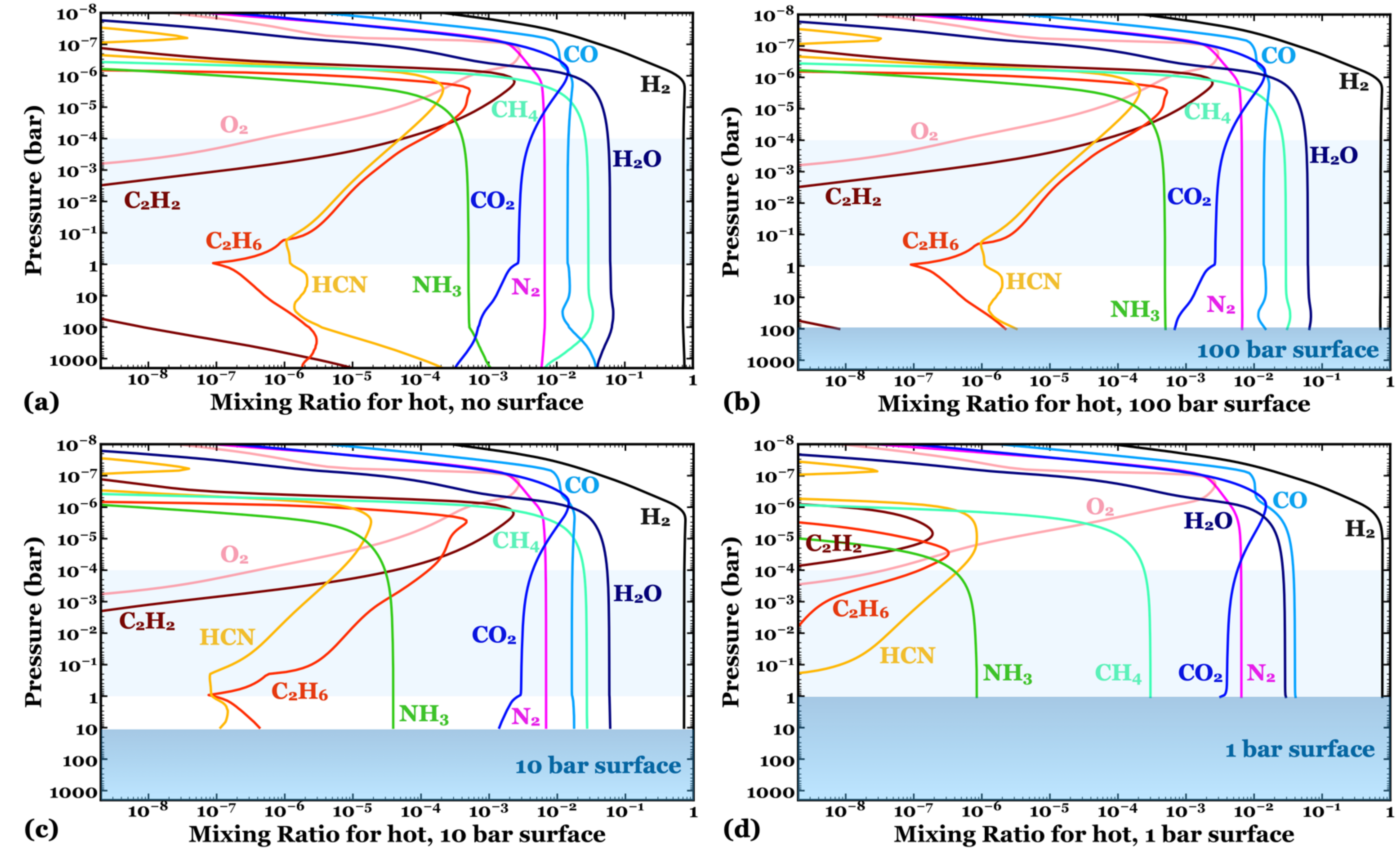}
\caption{Volume mixing ratio profiles for the main chemical species for a hotter sub-Neptune with K2-18b's physical parameters but four times its incident stellar flux (red P-T profile in Figure \ref{fig:TP}a) with T$_{\mathrm{int}}$ = 70 K, (a) without a surface (no-surface case) and with surface levels located at (b) 100-bar, (c) 10-bar, and (d) 1-bar levels.}
\label{fig:hot}
\end{figure}

In Figure \ref{fig:cold} and \ref{fig:hot}, we show the converged volume mixing ratio (VMR) profiles of the main chemical species for the nominal K2-18b and the hotter variant (both with T$_{\mathrm{int}}$ = 70 K). Different surface levels with different surface pressures and temperatures are shown to have significant impacts on the final chemical make-ups of the atmosphere. For the no-surface cases (Figure \ref{fig:cold}a and Figure \ref{fig:hot}a), large amounts of hydrocarbon and nitrile species are produced photochemically -- the main ones shown here include ethane (C$_2$H$_6$), acetylene (C$_2$H$_2$), and hydrogen cyanide (HCN) -- and other significant atmospheric species include H$_2$O, CH$_4$, CO, CO$_2$, NH$_3$, and N$_2$. Species such as H$_2$O and CH$_4$ have abundances close to their expected equilibrium values, whereas CO, CO$_2$, and N$_2$ are considerably enhanced and NH$_3$ is depleted in the upper atmosphere over equilibrium expectations, as a consequence of transport-induced quenching and (for the case of CO$_2$) related chemistry involving quenched species. The species for the 100-bar surface cases (Figure \ref{fig:cold}b and Figure \ref{fig:hot}b) have similar VMR profiles as the no-surface cases, as the surface temperatures at 100-bar ($>$ 1300 K) are already hot enough for thermochemistry to be effective, as if there were no surface. 

The 10-bar surface cases (Figure \ref{fig:cold}c and Figure \ref{fig:hot}c) start to deviate from the no-surface case, as the surface level is now well above the original deep quench point for the key nitrogen-containing species N$_2$, NH$_3$ and HCN, and chemical kinetics at these cooler temperatures no longer leads to efficient exchange of nitrogen between N$_2$ and NH$_3$. Some exchange of nitrogen between NH$_3$ and HCN still occurs, but the nitrogen lost from NH$_3$ photolysis largely ends up in the photochemically-stable N$_2$, so NH$_3$ becomes depleted, and then the coupled NH$_3$-CH$_4$ kinetics that produces HCN operates less efficiently. The 10-bar surface cases therefore exhibit decreased VMRs of NH$_3$ and HCN compared to the no-surface and 100-bar surface cases, whereas N$_2$ (with its strong triple N-N bond that makes it less susceptible to UV photolysis) exhibits an increase in its VMR. The 10-bar surface is also above the original quench point for CO-CH$_4$-H$_2$O, but surface temperatures are high enough that the photochemically produced complex hydrocarbons remain in equilibrium with CH$_4$ in the lower atmosphere, such that the photochemically produced C$_2$H$_x$ hydrocarbons and other species are recycled back to CH$_4$, preventing a significant depletion in atmospheric methane.

For the shallowest 1-bar surface cases tested in our model (Figure \ref{fig:cold}d and Figure \ref{fig:hot}d), the surface level is located well above the quench point for most species, leading atmospheric compositions to deviate significantly from the no-surface case. Instead of being tied to an equilibrium abundance at depth, the species vertical profiles represent a steady-state balance between photochemical production and loss and vertical transport. The column densities of photochemically-fragile H$_2$O, CH$_4$, and NH$_3$ drop significantly compared to the no-surface case, while those of the more photochemically-robust species CO, CO$_2$, and N$_2$ increase, as they become the major end products for the oxygen, carbon, and nitrogen released from the photochemistry of H$_2$O, CH$_4$, and NH$_3$. However, CO, CO$_2$, and N$_2$ and other key photochemical products can themselves be photolyzed by high-energy photons, and some of that oxygen, carbon, and nitrogen ends up back in H$_2$O, CH$_4$, and NH$_3$, so these original ``parent" species are not completely lost from the atmosphere. Instead, the vertical profiles of all the species evolve to reflect this complex chemical coupling and steady-state balance mentioned above. The resulting atmospheres end up being much ``cleaner" in terms of fewer complex, refractory, photochemical products compared to the no-surface cases. Most hydrocarbons and nitriles are depleted to VMR $<10^{-6}$ in this shallow 1-bar surface case.

\subsection{Sensitivity of trace species to different surface levels}

In order to find the species that can be used for probing the existence of surfaces, here in the section we examine the sensitivity of several key species to the existence of surfaces at different surface pressures.

\begin{table}
\caption{Species grouping based on their abundance response to the existence of surfaces.}
\label{table:group}
 \begin{center}
 \begin{tabular}{c c c}
 \toprule
Group & Characteristic & Species\\
\hline
1 & sensitive to all surface levels & NH$_3$, HCN\\
\hline
2 &only sensitive to 1-bar surface with decreased abundance & H$_2$O, CH$_4$, C$_2$H$_6$, C$_2$H$_2$\\
\hline
3 & only sensitive to 1-bar surface with increased abundance & CO, CO$_2$\\
\hline
4 & not sensitive to all surface levels & H$_2$, N$_2$, He\\
\toprule
\end{tabular}
\end{center}
\end{table}

\begin{figure}[h]
\centering
\includegraphics[width=\textwidth]{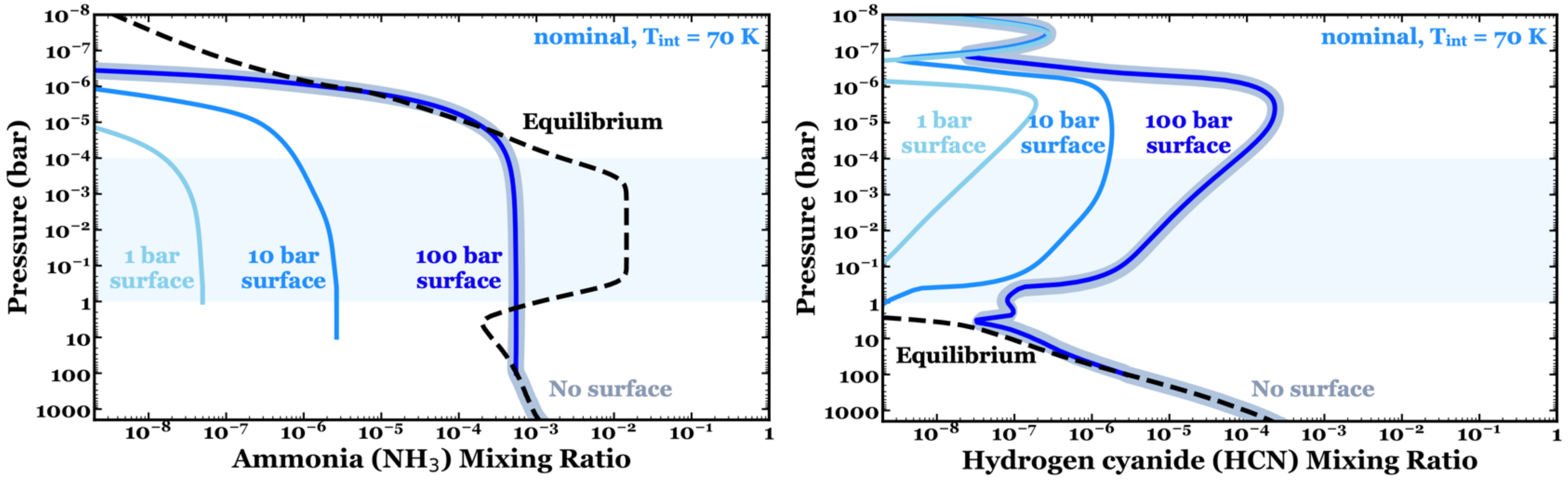}
\caption{The sensitivity of species in group 1 (NH$_3$ and HCN) to different surface levels for the nominal K2-18b case with T$_{\mathrm{int}}$ = 70 K. This group of species is sensitive to all surface levels with P$_{\mathrm{surf}}<$ 100 bar. The dash lines are the resulting VMR profiles using solely the thermochemical equilibrium model. The solid lines are for the full model with transport and photochemistry, with surfaces located at different levels (light blue: 1-bar surface, medium blue: 10-bar surface, dark blue: 100-bar surface, thick grey blue: no-surface). Note that the no surface case VMR profiles overlap with the profiles of the 100-bar surface case. The shaded blue region indicates the observable part of the atmosphere (10-4-1 bar).}
\label{fig:HCN_NH3}
\end{figure}

\begin{figure}[h]
\centering
\includegraphics[width=\textwidth]{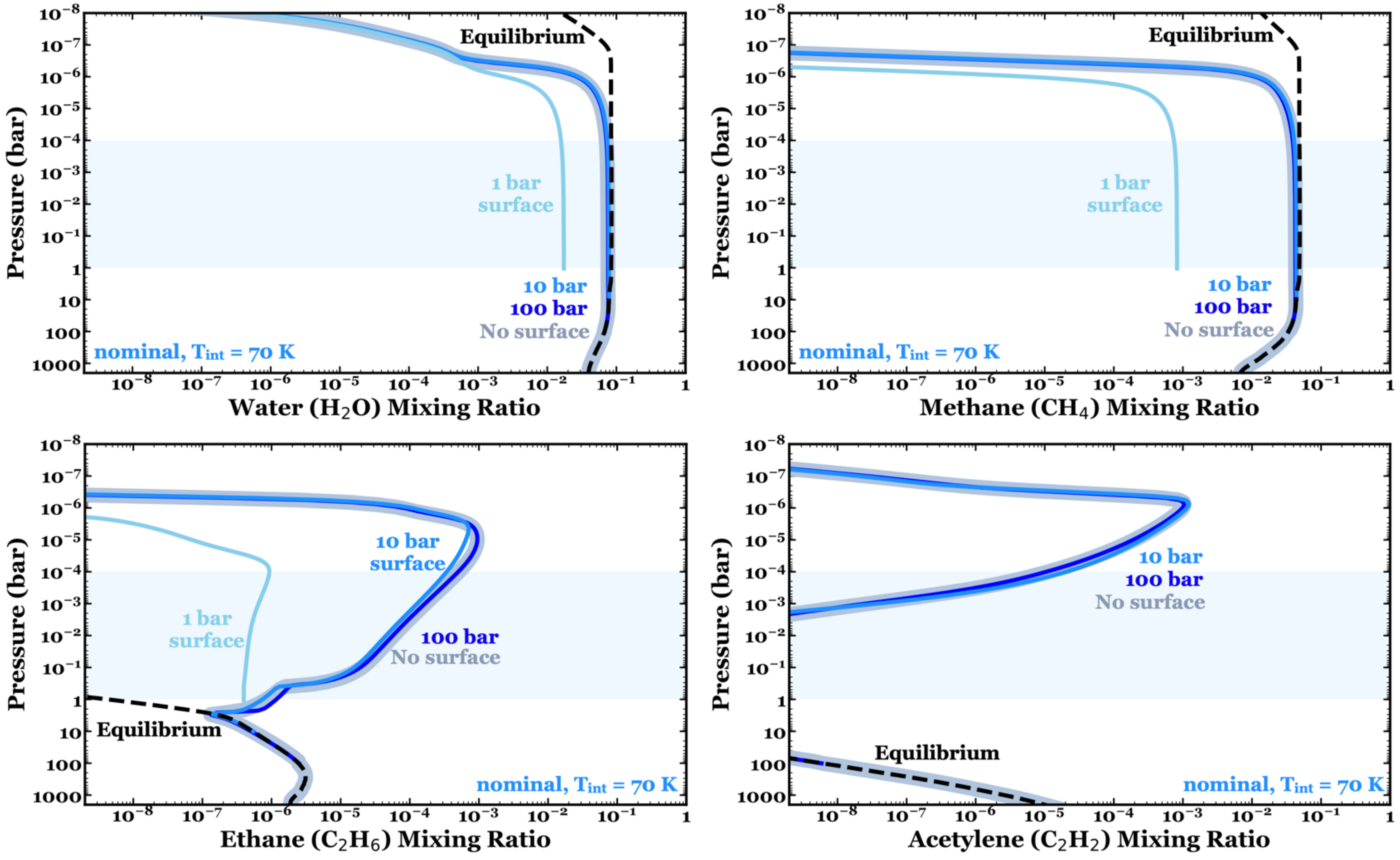}
\caption{The sensitivity of species in group 2 (H$_2$O, CH$_4$, C$_2$H$_6$, C$_2$H$_2$) to different surface levels for the nominal K2-18b case with T$_{\mathrm{int}}$ = 70 K. This group of species is sensitive only to the very shallow 1-bar surface, with decreased abundances compared to the no-surface case. The lines and the shade are labeled in the same way as in Figure \ref{fig:HCN_NH3}. Note that for H$_2$O and CH$_4$, the resulting abundance profiles for the 10-bar, 100-bar, and no-surface cases overlap each other. For C$_2$H$_6$, the 100-bar and no-surface results overlap each other. For C$_2$H$_2$, the 1-bar profile is not shown as it is below the lower bound of the mixing ratio shown in the plot ($<2\times10^{-9}$).}
\label{fig:CH4_H2O}
\end{figure}

\begin{figure}[h]
\centering
\includegraphics[width=\textwidth]{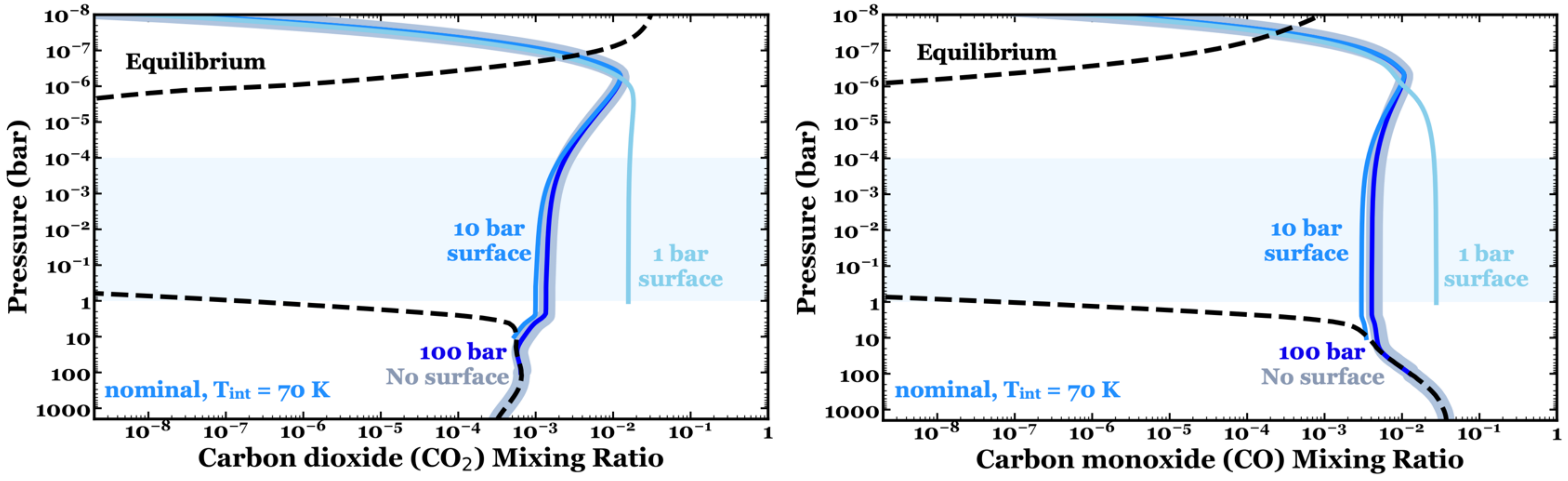}
\caption{The sensitivity of species in group 3 (CO$_2$ and CO) to different surface levels for the nominal K2-18b case with T$_{\mathrm{int}}$ = 70 K. This group of species is also sensitive to only very shallow 1-bar surfaces, similar to the group 2 species, but they have increased abundances compared to the no-surface case. The lines and the shade are labeled in the same way as in Figure \ref{fig:HCN_NH3}. Note that the resulting CO$_2$ and CO VMR profiles for the 100-bar and no-surface cases overlap each other.}
\label{fig:CO_CO2}
\end{figure}

\begin{figure}[h]
\centering
\includegraphics[width=\textwidth]{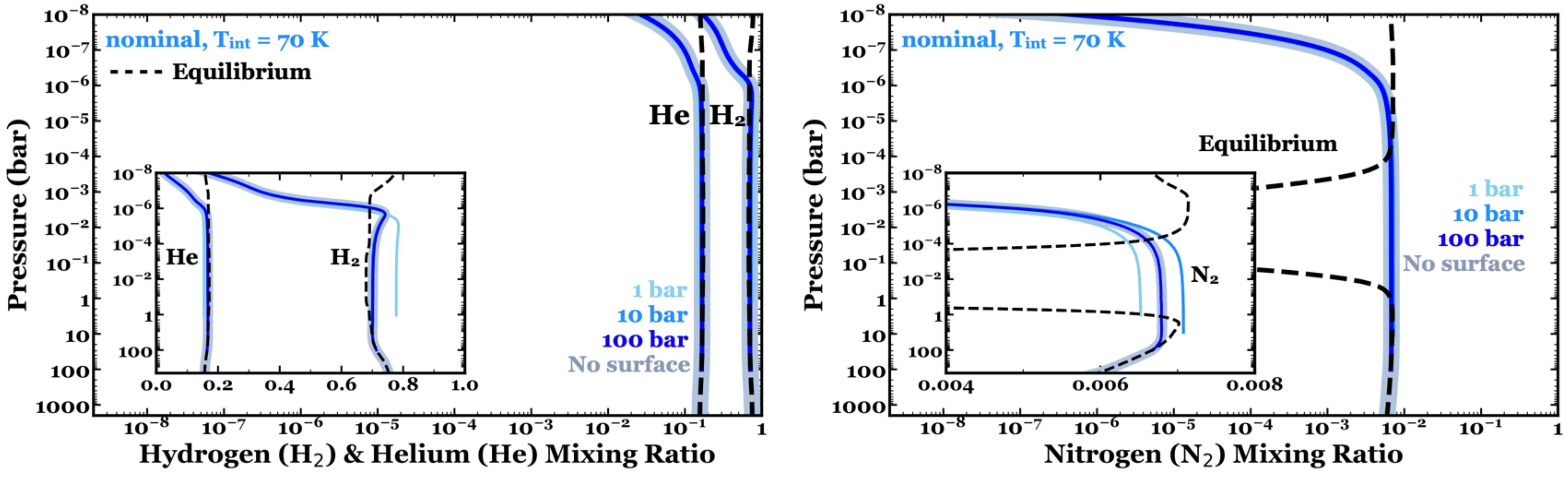}
\caption{The sensitivity of species in group 4 (H$_2$, He, N$_2$) to different surface levels for the nominal K2-18b case with T$_{\mathrm{int}}$ = 70 K. This group of species is not very sensitive to the existence of surfaces. The lines are labeled in the same way as in Figure \ref{fig:HCN_NH3}. The resulting H$_2$, He, and N$_2$ abundances for the 1-bar, 10-bar, 100-bar, and the no-surface cases overlap each other. The inset figure (plotted in linear scale) magnifies the differences in the VMR profiles of H$_2$, He, and N$_2$ for different surface level cases.}
\label{fig:H2_N2}
\end{figure}

Based on our results shown in Figure \ref{fig:cold} and \ref{fig:hot}, here we group the key chemical species into four categories (Table \ref{table:group} and Figures \ref{fig:HCN_NH3}--\ref{fig:H2_N2}), and each category of species is affected by the existence of a surface differently. The first category includes NH$_3$ and HCN, whose abundances are very sensitive to different surface levels (e.g., our 1-bar and 10-bar cases), such that their abundances decrease with decreasing surface pressure. The second category includes CH$_4$, other hydrocarbons (only C$_2$H$_6$, C$_2$H$_2$ are included in the table here), and H$_2$O; this category of species is only sensitive to very shallow cool surfaces (e.g., our 1-bar case), such that the mixing ratios of these species decrease in comparison to all other surface scenarios. The third category includes CO and CO$_2$, whose abundances are most sensitive to the very shallow cool 1-bar surface as well, but whose mixing ratios increase in the 1-bar surface case compared to all other surface scenarios. The last category includes species that are not very sensitive to the existence of surfaces; they are mostly non-polar or relatively inert species such as hydrogen, nitrogen, and helium (He). 

Figure \ref{fig:HCN_NH3} shows mixing ratio profiles for the species in the first category for the nominal K2-18b case (T$_{\mathrm{int}}$ = 70 K) with different surface levels at convergence. The thermochemical equilibrium profile of NH$_3$ has the highest VMR in the observable part of the atmospheres (10-4 to 1 bar). When vertical transport and photochemistry are included in the model, the NH$_3$ VMR for the no-surface case decreases from its expected equilibrium value by more than one order of magnitude in the observable region of the atmosphere, as a result of transport-induced quenching. At pressures less than $\sim$100 bar, the atmosphere is simply too cold for the thermochemical kinetics reactions to efficiently exchange nitrogen between different major species, and vertical transport is faster than the chemical reactions can maintain thermochemical equilibrium. However, at pressures greater than $\sim$100 bar, temperatures are high enough that the kinetic reactions dominate over transport, and thermochemical equilibrium can be kinetically maintained. As on Jupiter, any NH$_3$ lost by photolysis at higher regions of the atmosphere can be recycled at depth in these assumed hydrogen-dominated sub-Neptune atmospheres, and the NH$_3$ can be transported back throughout the atmosphere (see Figure \ref{fig:scheme}). Because the N$_2$-NH$_3$ quench point lies close to 100 bar, adding a surface at 100 bar does not cause significant changes in the NH$_3$ mixing ratio compared with the no-surface case; essentially, the 100-bar, $\sim$1300 K conditions near the surface allow the thermochemical kinetics reactions to efficiently recycle the NH$_3$ that is lost from photochemical processes higher up in the atmosphere, and NH$_3$ is not permanently lost. 

Ammonia is, however, very sensitive to surfaces located at lower pressures -- both the 10-bar surface case (T$_{\mathrm{surf}}\sim$ 1000 K) and the 1-bar surface case (T$_{\mathrm{surf}}\sim$ 600 K) exhibit a decrease in the NH$_3$ mixing ratio of $\sim$2 and $\sim$4 orders of magnitude, respectively, compared to the no-surface case. In the upper atmospheres of our shallow-surface models, some of NH$_3$ is irreversibly destroyed and converted mainly to N$_2$ and HCN, as the nitrogen in these two species forms strong triple-bonded structures that make them less chemically reactive and only able to be photolyzed by high-energy, extreme-ultraviolet photons. Note that hydrazine (N$_2$H$_4$), a major NH$_3$ photochemical product on Jupiter, is not synthesized in large quantities in these warmer sub-Neptune atmospheres, as numerous competitors for the NH$_2$ radicals exist to impede its production, and the produced N$_2$H$_4$ does not condense, which allows other chemical loss mechanisms to operate. 

In the shallow-surface models, without the thermochemistry that would occur deeper in the atmosphere, NH$_3$ cannot be recycled as efficiently, leading to the depletion of NH$_3$ in favor of N$_2$ compared to the no-surface and 100-bar surface cases. The fate of NH$_3$ on K2-18b is similar to that on Titan, where it is irreversibly converted to nitrogen, leading to a nitrogen-dominated atmosphere ($\sim$94-98\% N$_2$). However, because the atmosphere near our shallow-surface K2-18b cases is still relatively hot ($\sim$600 K for the 1-bar surface and $\sim$1000 K for the 10-bar surface) and because the background atmosphere is abundant in H$_2$, some NH$_3$ can still be recycled after it is photolyzed, either from NH$_2$ (through temperature-dependent reactions such as NH$_2$ + H$_2$ $\rightarrow$ NH$_3$ + H), or from N$_2$ \citep[through N$_2$ photolysis and subsequent reactions in the uppermost atmosphere or through schemes initiated by H-atom addition to N$_2$ at high pressures,][]{2010FaDi..147..103M, 2011ApJ...737...15M}. At convergence, NH$_3$ is not completely lost from the atmosphere. However, the lower the temperature and pressure at the surface, the less the NH$_3$ is able to be recycled. Being so sensitive to various surface levels less than 100 bar, NH$_3$ can be used as a key species to determine if an exoplanet has a surface or not. 

Chemical production of HCN is tied closely to NH$_3$ kinetics, so HCN is affected in a similar way as NH$_3$ to the presence of a surface at different pressure levels. As a result, HCN is also a key species for determining whether an exoplanet might possess a surface or not. Thermochemical equilibrium predicts minimal amounts of HCN in the observable atmospheres of K2-18b, while photochemistry and vertical transport significantly enhance the mixing ratio of HCN in those regions. The enhancement is largely the result of the coupled NH$_3$-CH$_4$ photochemistry, initiated by reactions such as CH$_3$ + NH$_2$ + M $\rightarrow$ CH$_3$NH$_2$ + M, where M is any third atmospheric molecule or atom, followed by reactions with atomic H to eventually form HCN \citep[see][]{2010FaDi..147..103M, 2011ApJ...737...15M}. The peak HCN mixing ratio at high altitudes in Figure \ref{fig:HCN_NH3} is a signature of this strong photochemical source. Similar to NH$_3$, the VMR profile of HCN is not affected by a 100-bar surface. This lack of sensitivity results from NH$_3$ being the main photochemical source of HCN -- because the ammonia profile is not affected, neither is HCN. Thermochemical kinetics at the surface conditions of 100 bar and $\sim$1300 K can fully recycle the HCN and NH$_3$ back to their larger equilibrium abundances at depth. The existence of a shallow, cool surface, however, drastically decreases the abundance of HCN by 1-2 orders of magnitude for the 10-bar surface case and 3-4 orders of magnitude for the 1-bar surface case compared to the no-surface case. Again, because HCN is the photochemical product of NH$_3$ and CH$_4$, when NH$_3$ is decimated in favor of N$_2$, the HCN abundance is also decreased. However, since HCN is expected to have a smaller column abundance than ammonia and thus is potentially harder to observe, it may not be as useful in predicting the existence of surfaces as NH$_3$.

Figure \ref{fig:CH4_H2O} shows the species in the second category, whose abundances in our models are only sensitive to very shallow, cool 1-bar surfaces. These species include H$_2$O, CH$_4$, and most heavier hydrocarbons (C$_x$H$_y$), of which only the most abundant ones, C$_2$H$_6$ and C$_2$H$_2$, are shown in Figure \ref{fig:CH4_H2O}. The thermochemical equilibrium model and the no-surface model produce similar amounts of H$_2$O and CH$_4$ in the observable part of the atmosphere due to the fact that these species are already the dominant oxygen and carbon carriers at the quench point in the no-surface model. Compared to the no-surface model, the very shallow 1-bar surface makes the mixing ratio of H$_2$O drop by a factor of $\sim$6 and the mixing ratio of CH$_4$ drop by almost 2 orders of magnitude. Similar to NH$_3$, some CH$_4$ is irreversibly converted to hydrocarbons photochemically in the upper atmosphere, with inefficient recycling of the complex hydrocarbons back to CH$_4$ at the surface conditions of 1 bar, 640 K. H$_2$O and CH$_4$ together are also photochemically converted to CO$_2$ and CO in the upper atmosphere via pathways involving O + CH$_3$ $\rightarrow$ H$_2$CO + H (among others), with no efficient thermochemical recycling at the 1-bar surface conditions, causing CO and CO$_2$ to eventually supplant CH$_4$ and H$_2$O as the dominant carbon and oxygen species in the atmosphere. Thus, we have the interesting result that a cool, shallow surface could lead to a reduced abundance of CH$_4$ and H$_2$O in favor of CO and CO$_2$ in sub-Neptune atmospheres, even when atmospheric metallicities are not high enough to favor CO and CO$_2$ in equilibrium.

Note that CH$_4$ and H$_2$O do not become as depleted as NH$_3$ because they are more efficiently recycled by thermochemistry and because their main photochemical products can be converted back to CH$_4$ and H$_2$O at lower temperatures than the main nitrogen-bearing photochemical products (HCN and N$_2$) can be converted back to NH$_3$. Photolysis of the complex hydrocarbons, for example, leads to pathways that can recycle methane in H$_2$-dominated atmospheres (see Moses et al., 2005). CO and CO$_2$ kinetics can also recycle the H$_2$O and CH$_4$ through reaction schemes such as H + CO$_2$ $\rightarrow$ OH + CO, followed by OH + H$_2$ $\rightarrow$ H$_2$O + H, or by:\\
H + CO + M $\rightarrow$ HCO + M\\
HCO + H$_2$ $\rightarrow$ H$_2$CO + H\\
H + H$_2$CO + M $\rightarrow$ CH$_2$OH + M\\
H + CH$_2$OH $\rightarrow$ OH + CH$_3$\\
OH + H$_2$ $\rightarrow$ H$_2$O + H\\
CH$_3$ + H$_2$ $\rightarrow$ CH$_4$ + H\\
--------------------------------------------\\
Net: CO + 3H$_2$ $\rightarrow$ H$_2$O + CH$_4$,

which occur more efficiently at higher temperatures and pressures but are still effective at atmospheric conditions relevant to the 1-bar surface cases. The final vertical profiles of the key species H$_2$O, CH$_4$, CO, and CO$_2$ represent a steady state between their coupled chemical production, loss, and vertical transport. Note that the case with CH$_4$ for K2-18b with a 1-bar surface is different from Titan (with a 1.5-bar surface), where CH$_4$ will be completely depleted, because 1) there is hardly any oxygen in Titan's atmosphere to permanently convert the carbon from CH$_4$ into CO and CO$_2$, and 2) Titan, with T$_{\mathrm{eq}}$ $\sim$ 80 K, is much colder compared to K2-18b (T$_{\mathrm{eq}}$ $\sim$ 255 K), and lots of the photochemically-produced hydrocarbons are condensable and would deposit on the surface of Titan to irrevocably deplete the atmospheric CH$_4$ over time \citep[e.g.,][]{2018SSRv..214..125A}, unless geological or biological source mechanisms are available to replenish the methane \citep[e.g.,][]{2006P&SS...54.1177A}. 

The warmer 10-bar and 100-bar surface (T$_{\mathrm{surf}}>$ 1000 K) models produce similar abundances of CH$_4$ and H$_2$O compared to the no-surface model. Under these surface pressure and temperature conditions, the photochemically-produced hydrocarbons and to some extent the CO and CO$_2$ are readily converted back to CH$_4$ and H$_2$O in the lower atmosphere and thus replenished, similar to the no-surface case. For the 10-bar surface case, CO$_2$ and CO are not completely in equilibrium with CH$_4$ and H$_2$O, but their mixing ratios have increased until a steady-state exchange of carbon and oxygen is maintained at a near-equilibrium situation, without a significant loss of H$_2$O and CH$_4$. In contrast, the lower atmosphere in the 1-bar surface case remains well out of thermochemical equilibrium.

For the hydrocarbons, the equilibrium model predicts minimal amounts in the observable part of the atmosphere. The no-surface model ends up with lots of hydrocarbons in the same region because of methane photochemistry \citep[e.g.,][]{2005JGRE..110.8001M}. The 100-bar and 10-bar surface results are similar to the no-surface case, as all the hydrocarbons (including methane) at these surface temperatures can be efficiently and thermochemically converted back and forth between each other. The 1-bar surface model ends up with significantly fewer hydrocarbons at convergence, predominantly because of the lower final abundance of CH$_4$. Actually, a lot of hydrocarbons are produced at the beginning of the 1-bar surface case ($<$ 0.1 Myr model runtime). However, because the photochemically-produced triple-bonded CO is less subject to photolysis and attack by atomic hydrogen compared to the photochemically-produced hydrocarbons, with time evolving, CH$_4$ and the photochemically produced hydrocarbons are eventually converted to CO and then to CO$_2$.

The species in group 2 are very sensitive to very shallow, cool surfaces (P$_{\mathrm{surf}}$ of a few bars or surface temperature T$_{\mathrm{surf}}<$ 800-900 K) but are not sensitive once the surface becomes too hot. Together with the group 1 species (which are sensitive to deeper, hotter surfaces), we can potentially identify not only the existence of surfaces but also the environmental conditions of the surfaces, such as pressure and temperature.

In Figure \ref{fig:CO_CO2}, we summarize VMR profiles for the species in the third category, CO$_2$ and CO -- both are also very sensitive to very shallow 1-bar surfaces but their abundances increase in comparison with the no-surface case. At equilibrium, CO$_2$ and CO have minimal abundances in the observable part of the atmosphere in this H$_2$-dominated, 100$\times$ metallicity atmosphere. With photochemistry and transport, the no-surface case produces significant amounts of CO$_2$ and CO. By adding a 1-bar surface, which significantly reduces thermochemistry in the deep hot part of the atmosphere, the oxygen in water and the carbon in methane are eventually converted to CO and CO$_2$, making them the dominant carrier of carbon and oxygen. As described in the previous category 2 species discussion, with CH$_4$ being inefficiently recycled from the photochemically-produced hydrocarbons at 1-bar surface conditions, CO outcompetes the hydrocarbons over the long term because it is less prone to photolysis or permanent destruction by reactions with atomic hydrogen. CO$_2$ also has increased abundance as it is readily produced from CO and H$_2$O kinetics. The 10-bar case is really interesting, as both CO$_2$ and CO are slightly less abundant compared to the no-surface case. The 10-bar surface is above the quench point of CO, and because the equilibrium abundance of CO increases with depth, the final quenched CO mixing ratio in the no-surface case is greater than the steady-state CO abundance in the 10-bar surface case. The decreased CO abundance for the 10-bar surface then leads to less kinetically produced CO$_2$. The 100-bar surface case is very similar to the no-surface case, as by putting a surface at this higher temperature and pressure, thermochemical reactions can fully recycle CH$_4$ and H$_2$O from CO$_2$ and CO.

Figure \ref{fig:H2_N2} includes species that are not very sensitive to the existence of surfaces, including H$_2$, He, and N$_2$. These species are all very abundant but are not readily observable in exoplanet atmospheres. Note that plotting the results case on a magnified linear scale makes it more obvious that the VMR profiles of H$_2$, He, and N$_2$ deviate slightly from the no-surface case. For the 1-bar surface model, H$_2$ has increased in VMR, and He and N$_2$ have decreased VMRs compared to the no-surface case (see the insets of Figure \ref{fig:H2_N2}). The increase in H$_2$ for the 1-bar case is caused by the loss of hydrogen from water and methane (Figure \ref{fig:CH4_H2O}) and the increased sequestration of the carbon and oxygen into CO and CO$_2$, which allows the released hydrogen to end up mostly in H$_2$, accompanied by a corresponding increase in the total number of molecules per unit volume at any pressure (at which point the mixing ratios are renormalized in the model). The increase in the atmospheric density caused by the increase in H$_2$ causes the VMRs of the other species to be correspondingly reduced. Thus, even though He is an inert species that does not participate in the chemical cycle, its mixing ratio is reduced compared to the no-surface case. The concentration of N$_2$ (number of molecules per unit volume) is increased compared to the no-surface case, but because the increase in H$_2$ concentration and total atmospheric density is more significant, the mixing ratio of N$_2$ in the 1-bar surface model ends up decreasing compared to the no-surface model. For the 10-bar surface model, the concentration of N$_2$ also increases compared to the no-surface case, but because H$_2$O and CH$_4$ are not lost as readily to H$_2$ at this surface level, the atmospheric density increase is minimal, thus, N$_2$ has increased mixing ratio compared to the no-surface case. 

\section{Observation strategy for identifying surfaces on sub-Neptunes}

Our photochemical and thermochemical kinetics modeling demonstrates that the mere presence of a surface on a sub-Neptune planet can circumvent thermochemical recycling mechanisms and alter the observable composition of its atmosphere, even if geological/biological processes are not actively providing sources or sinks of atmospheric gases. These results have important implications for future spectroscopic observations of sub-Neptunes, such as with JWST, ARIEL, and high-spectral-resolution ground-based observations. We find that trace species such as NH$_3$ and CH$_4$ can indeed be used as proxies for detecting exoplanet surfaces. In addition to NH$_3$ and CH$_4$, a few other species are found to be sensitive to the existence of surfaces, including HCN, complex hydrocarbons (here we choose the potentially observable C$_2$H$_2$), H$_2$O, CO, and CO$_2$. The existence of surfaces at depths where the atmosphere is relatively cool prevents the thermochemical kinetics reactions that would normally occur in the deep, hot part of atmosphere from recycling photochemically-consumed species and transporting them back to the upper atmosphere. Therefore, species that are less photochemically stable, such as NH$_3$, CH$_4$, and H$_2$O, and the photochemical products that depend on them (e.g., C$_x$H$_y$ hydrocarbons, HCN, CH$_3$CN, HC$_3$N, O$_2$) end up with lower abundances when a cool, shallow surface is present, whereas species that are harder to destroy by photochemistry (N$_2$, CO, CO$_2$) exhibit a corresponding increase in abundance. 

These trace species also have different responses to different atmospheric pressures/temperatures at the surface, as different species can approach their thermochemical equilibrium abundances under different conditions.. For conditions relevant to K2-18b in our nominal model, the CH$_4$-CO-H$_2$O quench point occurs at $\sim$30 bars and $\sim$1100 K, allowing efficient thermochemical kinetics exchange between CH$_4$, C$_x$H$_y$, hydrocarbons, H$_2$O, CO, and CO$_2$ under these or higher P-T conditions. Even if a surface were located at the 10-bar, $\sim$1000 K level, where the carbon and oxygen species are not fully equilibrated, exchange between CH$_4$ and C$_x$H$_y$ hydrocarbons is still very rapid, and exchange between H$_2$O, C-H bonded hydrocarbons, and CO/CO$_2$ is effective enough that most of the CH$_4$ and H$_2$O molecules that were destroyed by photochemistry end up being recycled in the lower atmosphere near the 10-bar surface and are transported back to the upper atmosphere, resulting in only minor changes in the abundance profiles of these species as a result of a surface at the 10-bar level. However, that recycling does not occur for a surface at 1 bar, $\sim$640 K. Meanwhile, the nitrogen species N$_2$ and NH$_3$ do not readily interact kinetically with each other until deeper and hotter conditions (N$_2$-NH$_3$ quench point near $\sim$90 bar, 1300 K), so these species -- and others such as HCN, whose chemical production and loss depends on NH$_3$ and/or N$_2$ -- are sensitive to the presence of a surface at pressures/temperature greater than 90 bar/1300 K.. 

\begin{figure}[h]
\centering
\includegraphics[width=\textwidth]{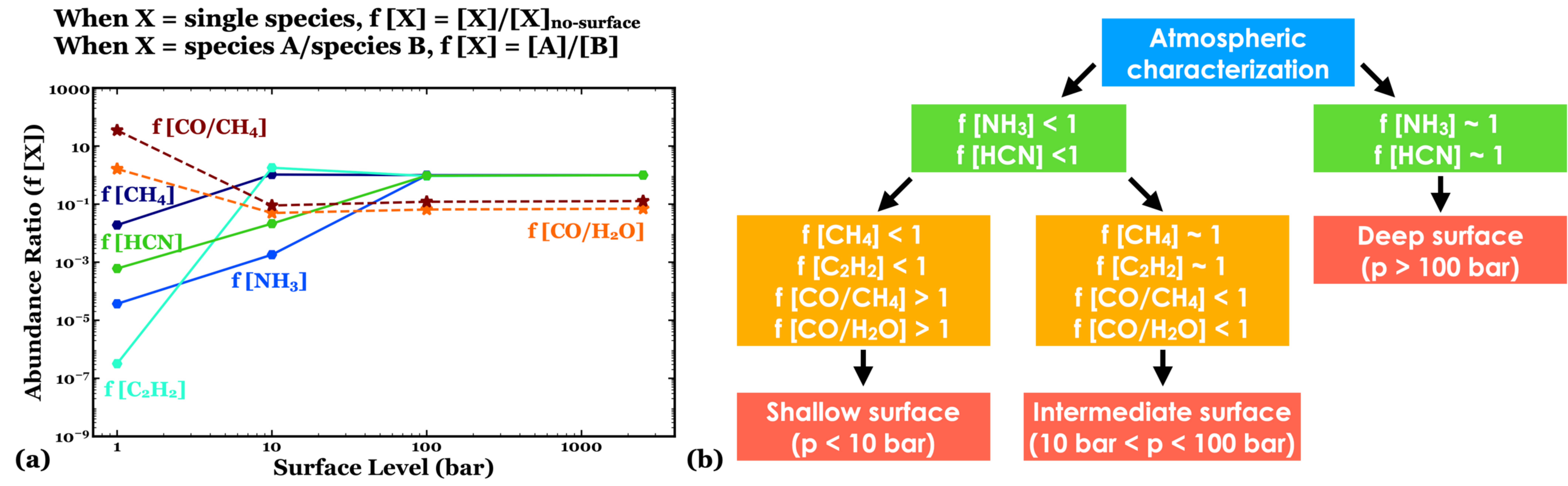}
\caption{(a) Effect of different surface levels on the abundance ratio of species for the nominal K2-18b with T$_{int}$ = 70 K at 10$^{-4}$ bar. (b) A flowchart of possible steps to identify the existence of surface and the surface level for a hydrogen-dominated exoplanet with properties similar to K2-18b. The process starts with the calculation of the abundances of species for the no-surface case, and then using a combination of the ratios between the observed abundances ([X]) and the no-surface abundances ([X]$_{\mathrm{no-surface}}$), f [X] = [X]/[X]$_{\mathrm{no-surface}}$, and the ratios between two species, f [X] = [A]/[B], we can predict whether this exoplanet has a deep surface (surface pressure, P$_{\mathrm{surf}}>$ 100 bar), an intermediate surface (10 bar $<$ P$_{\mathrm{surf}}<$ 100 bar), or a cold and shallow surface (P$_{\mathrm{surf}}<$ 10 bar).}
\label{fig:diagram}
\end{figure}

In Figure \ref{fig:diagram}, we propose a tentative flowchart to help observationally distinguish scenarios in which K2-18b or another sub-Neptune with a broadly similar instellation environment were to possess a surface located at various pressure levels; these observationally-based criteria depend on species abundances and their ratios. We first select criteria that would be satisfied for all our models cases and sensitivity tests with the varying P-T profiles and K$_{zz}$ profiles -- these criteria include the mixing ratios of NH$_3$, CH$_4$, HCN, and C$_2$H$_2$, and the abundance ratios between CO and CH$_4$, and CO and H$_2$O. For the single-species criteria, we use the abundance ratios between the observed mixing ratio ([X]) over the mixing ratio expected from the no-surface case ([X]$_{\mathrm{no-surface}}$):
\begin{equation}
f\ [X] = [X]/[X]_{\mathrm{no-surface}},\ \mathrm{when}\ X\ =\ \mathrm{single\ species}.
\end{equation}
The abundances of the no-surface case need first to be predicted from a photochemical-transport-thermochemistry coupled model and so are affected by uncertainties in the P-T profile and K$_{zz}$ profile, but such model-dependent comparisons represent a reasonable first attempt to quantify the effect of surfaces. We found that using thermochemical equilibrium abundances as the denominator in Equation (1) is hard to give consistent criteria, even though they are less subject to potential modeling uncertainties. This is because some species such as the heavier hydrocarbons, nitriles, CO, and CO$_2$ are expected to have negligible VMR in thermochemical equilibrium but are instead produced through disequilibrium chemistry. The abundance ratios between a few observable species pairs are also great criteria for revealing the existence of cool, shallow surfaces (as shown in Figure \ref{fig:diagram}):
\begin{equation}
f\ [X] = [A]/[B], \mathrm{when}\ X = \mathrm{species}\ A/\mathrm{species}\ B.
\end{equation}
These criteria include the ratios between CO and CH$_4$ (f [CO/CH$_4$]), and CO and H$_2$O (f [CO/H$_2$O]). For a shallow surface (P $<$ 10 bar, T $<$ 1000 K), the carbon and oxygen preferentially reside in CO over CH$_4$ and H$_2$O, leading to f [CO/CH$_4$] and f [CO/H$_2$O] being less than unity. For a deeper surface (P $\geq$ 10 bar, T $\geq$ 1000 K) under relevant K2-18b conditions, carbon and oxygen preferentially stay in CH$_4$ and H$_2$O over CO.

Overall, f [NH$_3$] and f [HCN] can be sensitive indicators of the existence of surfaces on sub-Neptunes, with values closer to unity for f [NH$_3$] and f [HCN] indicating no surface or a deep surface with a pressure level of at least 100 bar. If f [NH$_3$] and f [HCN] are much smaller than 1, then we can use other criteria to distinguish whether there is a cool, shallow surface or an intermediate surface, including f [CH$_4$], f [C$_2$H$_2$], f[CO/CH$_4$], f [CO/H$_2$O]. If f [CH$_4$] and f [C$_2$H$_2$] are less than 1 and f [CO/CH$_4$] and f [CO/H$_2$O] are larger than 1, a cool, shallow surface may be indicated. Larger-than-unity f [CH$_4$] and f [C$_2$H$_2$] and smaller-than-unity f [CO/CH$_4$] and f [CO/H$_2$O] would point to an intermediate surface. For the case where HCN and C$_2$H$_2$ are not observable at the mixing ratios predicted for the no-surface model, due to limited instrument sensitivity or wavelength coverage, then the other criteria that include the more-abundant species (CH$_4$, NH$_3$, CO, H$_2$O) should be sufficient for determining the existence of surface and constraining surface conditions.

\begin{figure}[h]
\centering
\includegraphics[width=\textwidth]{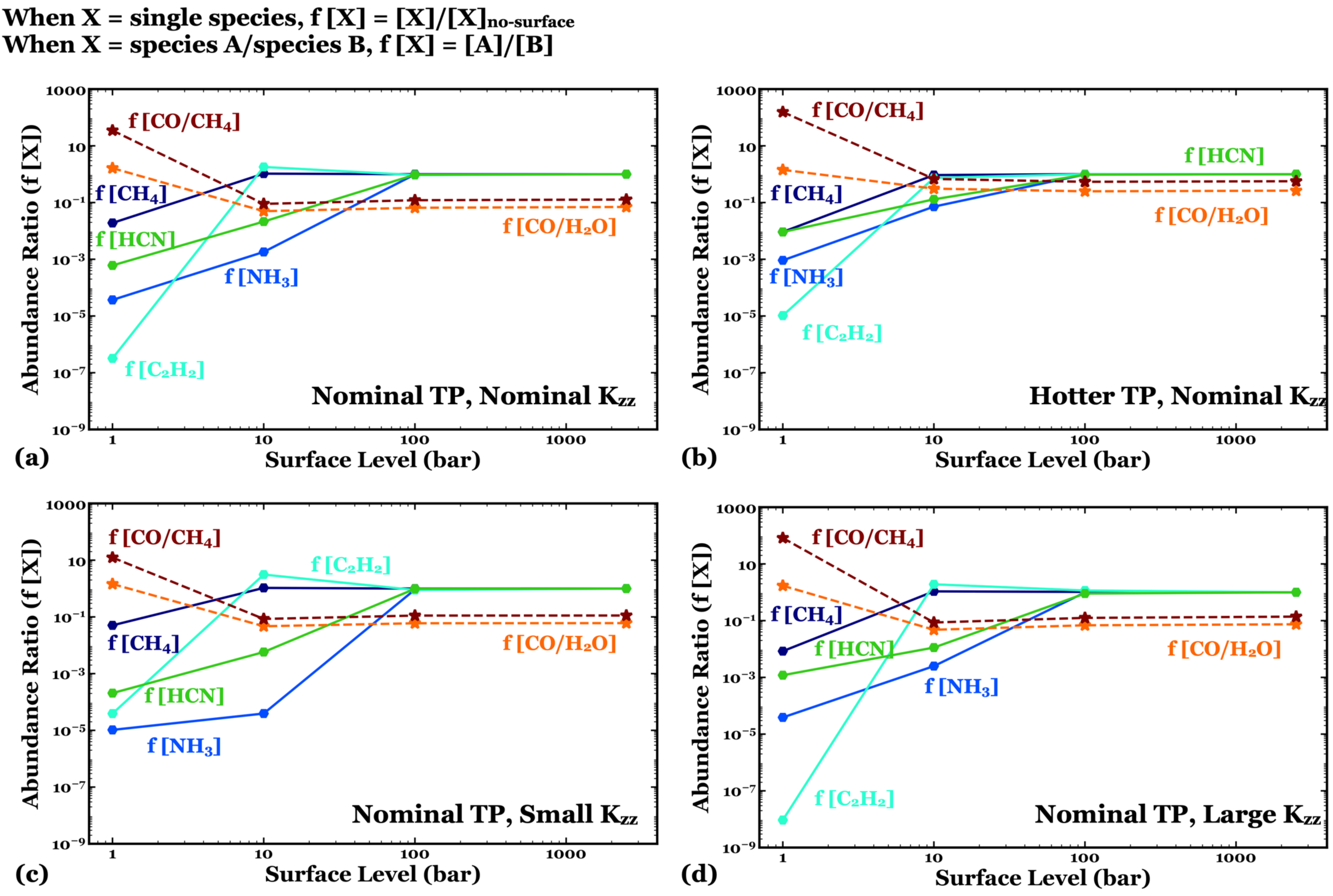}
\caption{Sensitivity of abundance ratio of species (at 10$^{-4}$ bar) with different surface levels for: (a) the nominal K2-18b with T$_{int}$ = 70 K and the nominal K$_{zz}$ profile, (b) the hotter K2-18b variant with T$_{int}$ = 70 K at 10$^{-4}$ bar, (c) the nominal K2-18b with T$_{int}$ = 70 K and a small K$_{zz}$ profile (nominal profile, see Figure 1, divided by three), (d) the nominal K2-18b with T$_{int}$ = 70 K and a large K$_{zz}$ profile (nominal profile multiplied by three).}
\label{fig:diagram_sensitivity}
\end{figure}

We also tested the robustness of our results to different model assumptions, by changing planetary parameters such as the atmospheric thermal structure and strength of vertical mixing. Figure 10 shows the sensitivity of the abundance ratios versus surface levels (Figure \ref{fig:diagram}) to different planetary parameters, including the T-P profile and the assumed K$_{zz}$ profiles. The results for a hotter variant of K2-18b is shown in Figure \ref{fig:diagram_sensitivity}b and the results for a smaller and a larger K$_{zz}$ profiles are shown in Figure \ref{fig:diagram_sensitivity}c and \ref{fig:diagram_sensitivity}d. Overall, our main qualitative conclusions in Section 3.2 and the abundance ratios we selected for Figure 9 remain true for these different planetary parameters. For a more detailed explanation of the effect of each parameter (T$_{int}$, P-T profile, and K$_{zz}$ profile) on each species, please refer to the Appendix section.

The application of our results to K2-18b today is hindered by the diverse range of retrieval results for the current atmospheric compositions of the planet \citep{2019ApJ...887L..14B, 2020arXiv201110424B, 2020ApJ...891L...7M, 2020ApJ...898...44S, 2021A&A...646A..15B, 2021a&a201111553C}. For instance, a CH$_4$ mole fraction around 0.03-0.1 was retrieved in \citet{2020arXiv201110424B} and \citet{2021a&a201111553C}, potentially indicating a surface deeper than 1 bar (see Figure \ref{fig:CH4_H2O}); while CH$_4$ was retrieved to be $<6.4\times10^{-4}$ in \citet{2020ApJ...898...44S}, which potentially indicates a very shallow 1-bar surface. Given the current data signal-to-noise ratio and wavelength coverage, we are unable to use the current retrieved results to determine the existence of a surface on K2-18b. We need the precise retrieval abundances of NH$_3$, CH$_4$, CO, H$_2$O through future JWST, ARIEL, and ground-based observations with higher sign-to-noise ratio and larger wavelength coverage to perform the surface determination in Figure \ref{fig:diagram}.

\section{Discussion}

Our results also have a few other implications. Being able to identify the existence of surfaces and the surface locations on intermediate-sized exoplanets could help us verify or distinguish planet formation theories. For example, if exoplanets on the high end of the radius gap valley and above tend to have no surfaces or deep surfaces, that would support the mass-loss theories \citep{2013ApJ...775..105O, 2013ApJ...776....2L, 2018MNRAS.476..759G}, while if they tend to have intermediate or shallow surfaces (the gas-liquid interface), that would instead favor the water-rich worlds theories \citep{2019PNAS..116.9723Z, 2020ApJ...896L..22M}.

We note that for exoplanets with shallow, cool surfaces (the 1-bar model), the H$_2$O abundance is decreased $\sim$6 times compared to the deep surface and the 10-bar surface model. Thus, for exoplanets with shallow, cool surfaces, observing solely the abundance of H$_2$O may not provide a good indication of their actual atmospheric metallicities, as a substantial percentage of the available oxygen is tied up in other species. Observations that constrain the mixing ratios of H$_2$O, CO, and CO$_2$ in tandem would provide a better metallicity estimate. For exoplanets with shallow, cool surfaces, the atmospheres also evolve to have fewer reduced species (H$_2$O, CH$_4$) and a greater abundance of oxidized species (CO, CO$_2$) and are also ``cleaner" (fewer photochemically-produced hydrocarbons and nitriles), while the atmospheres contain more reduced species and are ``dirtier" (with lots of hydrocarbons and nitriles) if the surface is deeper. This difference may also impact photochemical haze formation \citep{2018NatAs...2..303H, 2018AJ....156...38H, 2018ApJ...856L...3H, 2019ECS.....3...39H, 2020NatAs...4..986H, 2020PSJ.....1...51H, 2020PSJ.....1...17M, 2021PSJ.....2....2V, 2021natas...55..465Y}.

The chemical composition of exoplanets with surfaces tend to deviate much more from thermochemical equilibrium compared to exoplanets without surfaces. This suggests that thermochemical equilibrium models may provide a reasonable starting point to predict atmospheric compositions for large, hot exoplanets with no surfaces \citep[e.g.,][]{1999ApJ...512..843B, 2002Icar..155..393L, 2007ApJS..168..140S, 2008ApJ...678.1419F, 2010ApJ...716.1060V}, but photochemical models are likely needed to predict the actual atmospheric compositions for smaller exoplanets that potentially have surfaces.

Our study seeks to understand the effect of planetary surfaces on the compositional evolution of exoplanet atmospheres. Using K2-18b as an example planet is just the start point of this pioneering study. More modeling is needed to cover a larger range of planetary parameter space (including different planetary physical and orbital parameters, different stellar parameters, different assumed atmospheric metallicities, different bulk elemental abundances, the addition of possible interior outgassing and/or sequestration of species at the surface, the inclusion of condensation, and so on). A wider range of models, in combination with atmospheric characterization data, could be used to predict the existence and the approximate conditions of surfaces for sub-Neptunes. For example, our current models predict that if some fraction of warm sub-Neptunes (i.e., with radii above the photoevaporation valley, such that they still possess some H/He) have shallow, cool surfaces, then the population as a whole would appear to have less NH$_3$ and HCN and a greater (CO + CO$_2$)/CH$_4$ ratio than their inferred atmospheric metallicities and effective temperatures would predict for the assumption of deep atmospheres and effective thermochemical recycling at depth. If this model prediction holds true for a variety of planetary parameters and scenarios, comparisons with observations might reveal whether shallow surfaces are statistically probable in this population or not. Other observational data can also help to test our surface predictions. For example, \citet{2019Natur.573...87K} detected the presence of a surface (or the lack of a thick atmosphere) on a close-in super-Earth LHS 3844b (1.3R$_{\mathrm{Earth}}$, 11 hr orbit), using thermal phase curve data. In addition to thermal phase curves, secondary eclipse observations may be able to distinguish the presence of a thick vs a thin atmosphere \citep{2019ApJ...886..141M, 2019ApJ...886..140K, 2019ApJ...886..142M, 2020ApJ...893..161M}.

It is interesting to note that methane has been found to be unexpectedly depleted on a few warm Neptune-class planets and sub-Neptunes for which we have good spectroscopic data: GJ 436b \citep{2010Natur.464.1161S, 2012ApJ...755....9S, 2011ApJ...735...27K, 2014Natur.505...66K, 2011ApJ...729...41M, 2011ApJ...738...32L, 2014A&A...572A..73L}, WASP-107b \citep{2018ApJ...858L...6K}, and GJ 3470b \citep{2019NatAs...3..813B}. Significant photochemical depletion of CH$_4$ is not expected under the relevant conditions for these planets in the absence of a surface, and although high atmospheric metallicity and/or high interior temperatures have been proposed as a potential explanation \citep[e.g.,][]{2013ApJ...777...34M, 2014ApJ...781...68A, 2017AJ....153...86M, 2019NatAs...3..813B, 2020AJ....160..288F}, our work is at least suggestive that novel pathways for CH$_4$ loss should be investigated for these planets as well.

Our currently adopted T-P profiles for the different surface locations are simply the truncated versions of the P-T profiles of the no-surface case at different surface pressure levels. Incorporating the effect of a surface in future radiative-convective modeling would provide more realistic thermal profiles near the surface for planets with surfaces. In our work here, the two end-member choices for T$_{\mathrm{int}}$ (0 and 70 K) certainly encompass the widest range of possibilities of K2-18b deep atmosphere temperatures, but more specific future predictions could utilize a more refined surface condition.

Our current chemical model is based on one developed for hotter giant planets and considers only neutral reactions between H, O, C, and N species; real planetary atmospheres can be more complicated. Future models investigating the topic could include more comprehensive reaction lists to help better confirm and constrain these results. For example, this work could also be adapted to other photochemically fragile species such as PH$_3$ and H$_2$S, both of which are detectable by JWST \citep{2017ApJ...850..199W}. Both sulfur and phosphorus species can significantly affect photochemistry in exoplanet atmospheres \citep{2009ApJ...701L..20Z, 2020NatAs...4..986H}, but photochemical reaction kinetics involving sulfur and phosphine species are not well constrained. However, both PH$_3$ and H$_2$S have shorter photochemical lifetimes than CH$_4$ and NH$_3$, thus they are likely more depleted in exoplanets with surfaces compared to the ones with deep/no surfaces. The inclusion of ion chemistry into the model would also be important. Ion chemistry in Titan's upper atmosphere is responsible for the production of refractory molecules with large molecular masses up to 10,000 m/z \citep{2007GeoRL..3422103C}; thus, the inclusion of ion chemistry could lead to additional sinks for carbon, nitrogen, and oxygen to a cool surface through the deposition of refractory aerosols (see arrow with question mark in Figure \ref{fig:scheme}).

Perhaps most importantly, our current model assumes zero flux of all species at both the upper and the lower boundary. Considering hydrogen escape at the top of the atmosphere may more realistically simulate the situation occurring in irradiated sub-Neptune atmospheres and may further deplete NH$_3$, CH$_4$, and H$_2$O if hydrogen is preferentially lost at very old planet ages \citep{2020ApJ...896...48M}. Including downward and upward fluxes at the lower boundary for certain gases to represent atmosphere-surface-interior interactions in future models would also more realistically simulate atmospheric evolution for terrestrial planets. Once an exoplanet has a surface, geological processes such as volcanic outgassing and deep-sea serpentine hydrothermal vents could release gases into the atmospheres to replenish the depleted species and participate in atmospheric chemistry \citep[e.g.,][]{2008ApJ...685.1237E, 2014E&PSL.403..307G, 2020PSJ.....1...58W, 2021natas...sdsY}, creating upward fluxes. For example, the current abundance of CH$_4$ on Titan is likely a result of interior outgassing \citep[e.g.,][]{2006Natur.440...61T}. Conversely, other species might be lost once they encounter the surface, as a result of chemical weathering, dissolution and reactions with water or magma oceans \citep[e.g.,][]{2016ApJ...829...63S, 2018ApJ...854...21C, 2019ApJ...887L..33K, 2020ApJ...891..111K}, condensation \citep[e.g.][]{2019ApJ...887L..14B}, or other surface-atmosphere interactions, creating downward fluxes. Thus, if the observed abundance ratios of species contradict each other using the flowchart in Figure 9, this could indicate that atmosphere-surface-interior exchanges are at work.

Note that most results shown here are converged results after billions of years and may not be realistic for a young exoplanet. For example, the 1-bar surface model initially produces large amounts of hydrocarbons and nitriles through photochemistry in the first 0.1 Myr years, however, with time evolving, the photochemically-formed nitriles and hydrocarbons are themselves photolyzed or kinetically destroyed and are eventually converted to nitrogen, carbon dioxide, and carbon monoxide. For extremely young exoplanets, it is possible that we are seeing a snapshot of an evolving atmosphere, whose species abundances may not reflect a truly steady-state atmosphere.

\section{Conclusion}
Our kinetics-transport model for K2-18b demonstrates that the presence of a surface can significantly alter the atmospheric abundances in a hydrogen-dominated sub-Neptune atmosphere. We identify a few potentially observable trace species that are affected by the inclusion of a cool, shallow surface: NH$_3$, HCN, CH$_4$, C$_2$H$_2$, H$_2$O, CO, and CO$_2$. The change in abundance of these species with an inclusion of a surface is due to the fact that thermochemical kinetics, which occurs efficiently in the deep, hot part of the atmosphere, is inhibited at lower temperatures and pressures. The presence of a cool surface at low atmospheric pressures can shut down or significantly impede thermochemical recycling. Thus, photochemically-fragile species (NH$_3$, HCN, CH$_4$, C$_2$H$_2$, H$_2$O) are destroyed over time, with no recycling from the deep atmosphere, and photochemically-stable species (CO, CO$_2$) survive and build up in the atmosphere. 

The atmospheric abundances of these species are also affected differently by different surface conditions, because each species can kinetically approach thermochemical equilibrium at different pressure/temperature points. Among all the species, the abundances of CH$_4$, C$_2$H$_2$, H$_2$O, CO, and CO are only affected by cool, low-pressure surfaces, as their equilibrium point is relatively shallow ($\sim$30 bar, $\sim$1100 K for expected K2-18b thermal conditions). The abundances of NH$_3$ and HCN are affected by deeper surfaces because they have deeper equilibrium points ($\sim$90 bar, $\sim$1300 K for K2-18b thermal conditions).

We also identify combinations of these species that can serve as proxies for identifying a surface and evaluating the approximate surface conditions of K2-18b or other similar sub-Neptunes. We expect this framework to be applied to other small observable exoplanets in the future.

\section{Acknowledgements}
X. Yu is supported by the 51 Pegasi b Postdoctoral Fellowship. J. Moses acknowledges support from NASA grant 80NSSC20K0462. J. Fortney acknowledges support from NASA grant 80NSSC19K0446. X. Zhang acknowledges support from NASA grant 80NSSC19K0791. The photochemical model results are available at https://doi.org/10.7291/D1338M.
\section{Appendix}

\subsection{Sensitivity to planetary parameters}

In the previous section, we identified a few species that can be potentially used as proxies for determining the existence of surfaces and surface conditions. Here we test the robustness of our results to different model assumptions, by changing planetary parameters such as the atmospheric thermal structure and strength of vertical mixing. Overall, our main qualitative conclusions in Section 3.2 remain true for species with various changed planetary parameters. In the following paragraphs, we will discuss the detailed effects of individual planetary parameters on the abundance profiles of key species listed in Table 1.

\begin{figure}[h]
\centering
\includegraphics[width=\textwidth]{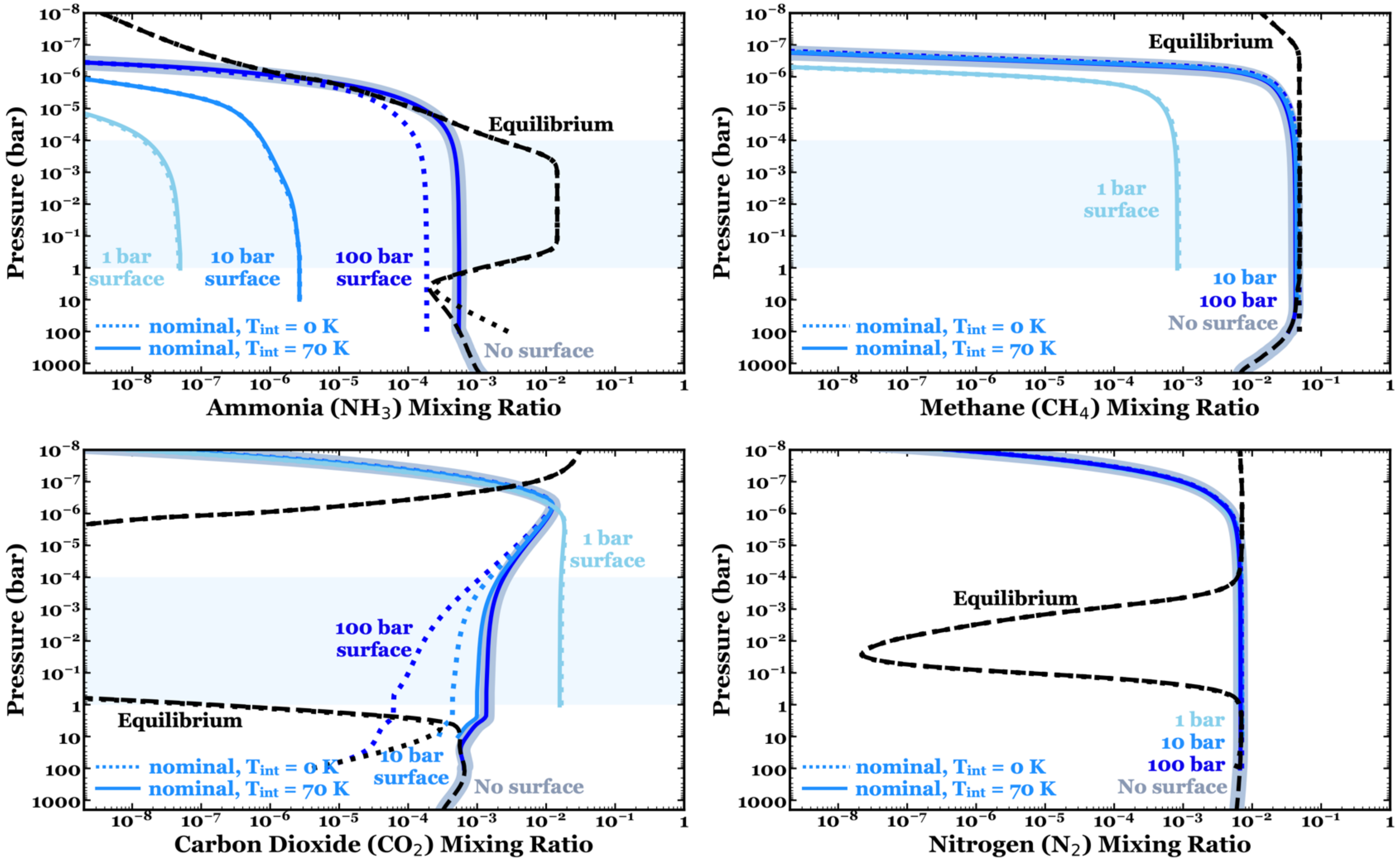}
\caption{The sensitivity of species abundances to P-T profiles with different intrinsic temperature (T$_{\mathrm{int}}$ = 0 K and T$_{\mathrm{int}}$ = 70 K) for the nominal K2-18b case. For the sensitivity study plots, we choose one representative species from each group (NH$_3$ from group 1, CH$_4$ from group 2, CO$_2$ from group 3, and N$_2$ from group 4) to exemplify the differences. Lines are colored in the same way as in Figure \ref{fig:HCN_NH3}. The solid lines are for the results with a higher internal heat flux, such that T$_{\mathrm{int}}$ = 70 K (the nominal case), and the dotted lines are for an internal heat flux of zero, or T$_{\mathrm{int}}$ = 0 K. Note that the equilibrium profiles for the T$_{\mathrm{int}}$ = 0 K case (dotted black lines) overlap with the T$_{\mathrm{int}}$ = 70 K case (dashed black lines) in the upper atmosphere (p $<\sim$ 10 bar).}
\label{fig:Tint}
\end{figure}

We first examine the effect of temperature in the deep atmosphere by testing two end-member cases of the internal heat flux, as represented by an intrinsic temperature T$_{\mathrm{int}}$ of 0 K and 70 K (the latter being our nominal case). As shown in Figure \ref{fig:TP}, changing T$_{\mathrm{int}}$ causes the P-T profile to vary in the deep atmosphere at pressures greater than a few bars, but not at lower pressures. Thus, the 1-bar surface case results are unaltered by the change of T$_{\mathrm{int}}$, as can be seen in Figure \ref{fig:Tint}. For the 10-bar surface case, the surface temperature is only $\sim$50 K higher for the T$_{\mathrm{int}}$ = 70 K case compared to the T$_{\mathrm{int}}$ = 0 K case; thus, most species (species in group 1, 2, 4) have similar VMR profiles. The only exceptions are the species in group 3, CO and CO$_2$, whose steady-state kinetics are quite sensitive to temperature and pressure. Although CH$_4$ and more complex hydrocarbons interact readily with each other at the P-T conditions relevant to the 10-bar surface for both T$_{\mathrm{int}}$ cases, the carbon species as a whole are not in thermochemical equilibrium at any pressure in these atmospheres, and full exchange between CO and CH$_4$ at depth, for example, is inhibited. However, kinetic reactions still occur, allowing the photochemically-produced CO and CO$_2$ to be destroyed throughout the atmospheric column, including near the surface. The cooler conditions at depth in the T$_{\mathrm{int}}$ = 0 K model favor the kinetic maintenance of CH$_4$ over CO, so there is less CO in the T$_{\mathrm{int}}$ = 0 K model than the T$_{\mathrm{int}}$ = 70 K model, and CH$_4$ is more readily recycled. The decreased CO abundance in the T$_{\mathrm{int}}$ = 0 K model leads to decreased net chemical production of CO$_2$, thus its abundance is also decreased. 

For the 100-bar surface case, the carbon and oxygen species are able to eventually follow their equilibrium abundances at depth in the higher-temperature T$_{\mathrm{int}}$ = 70 K model, but these species remain out of equilibrium in the cooler T$_{\mathrm{int}}$ = 0 K model. The kinetics at high pressures in these H$_2$-rich atmospheres increasingly favors CH$_4$ over CO, especially at colder temperatures, so VMRs of group 3 species, CO and CO$_2$, become significantly depleted in the colder T$_{\mathrm{int}}$ = 0 K, 100-bar model in comparison to the T$_{\mathrm{int}}$ = 70 K, 100-bar model. For species in group 2 (CH$_4$ shown as an example in Figure \ref{fig:Tint}), the 100-bar surface temperature is hot enough for thermochemistry to fully recycle the hydrocarbon photochemical products back to methane and transport CH$_4$ back to the upper atmosphere. Thus, the abundances of group 2 species show little sensitivity to T$_{\mathrm{int}}$ when the surface is placed at 100 bar. 

Similarly, the nitrogen species never achieve thermochemical equilibrium at the surface conditions relevant to the T$_{\mathrm{int}}$ = 0 K, 100-bar model, but they come very close to it in the T$_{\mathrm{int}}$ = 70 K model, 100-bar model. This causes the species in group 1 (NH$_3$ shown as an example in Figure \ref{fig:Tint}) to exhibit sensitivity to T$_{\mathrm{int}}$ in the 100-bar models. Note that at 100 bar, the surface temperature is $\sim$350 K lower in the T$_{\mathrm{int}}$ = 0 K case in comparison to the T$_{\mathrm{int}}$ = 70 K case. Thermochemical kinetics in the lower atmosphere cannot fully recycle the NH$_3$ that was lost photochemically in the upper atmosphere in the T$_{\mathrm{int}}$ = 0 K model, in contrast to the warmer T$_{\mathrm{int}}$ = 70 K model, leading to lower NH$_3$ abundances in the T$_{\mathrm{int}}$ = 0 K case. Because HCN is generated predominantly from ammonia photochemistry, the decreased NH$_3$ abundance in the T$_{\mathrm{int}}$ = 0 K model leads to less HCN, as well.

\begin{figure}[h]
\centering
\includegraphics[width=\textwidth]{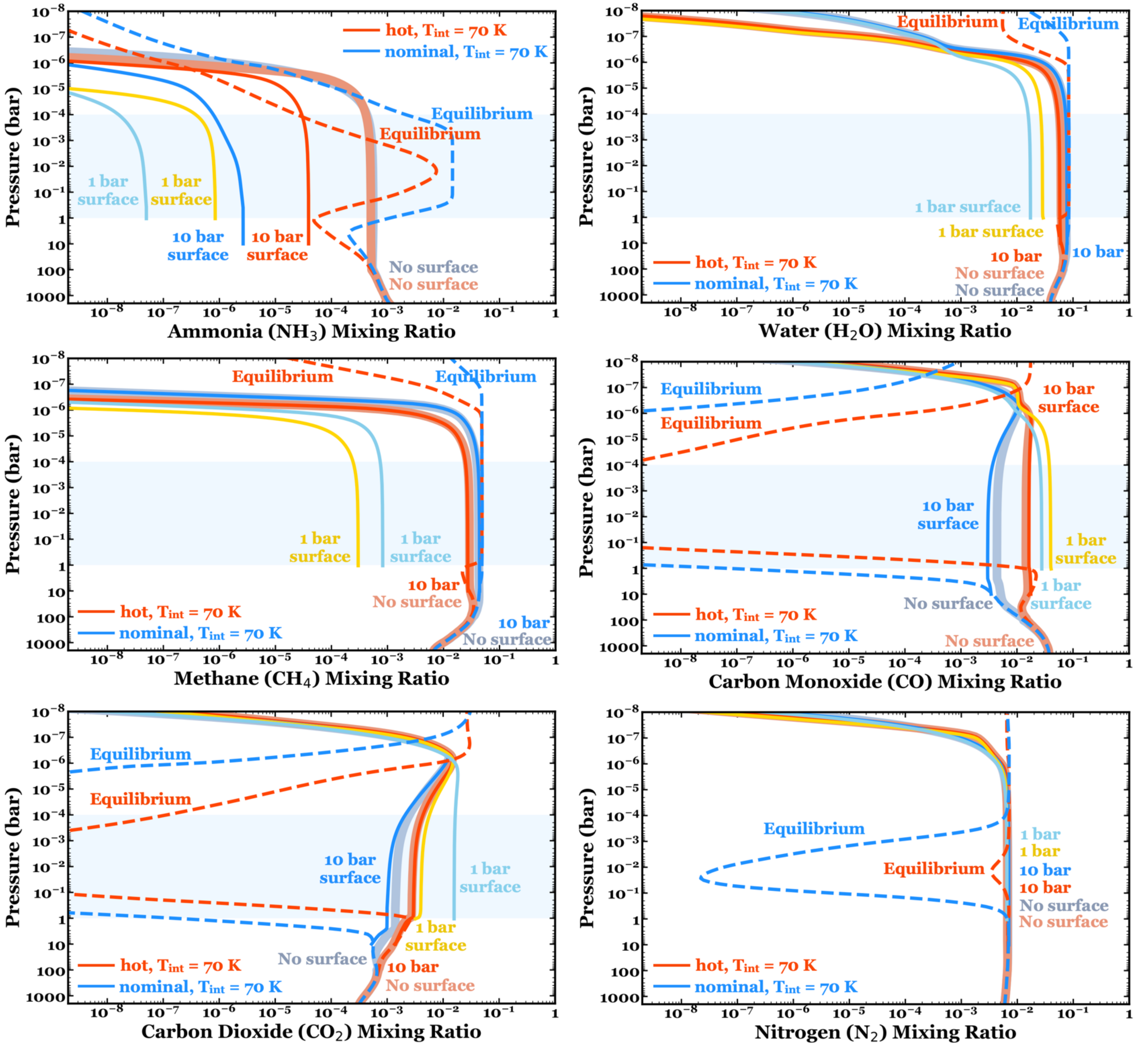}
\caption{The sensitivity of species abundances to orbital distance, where the ``hot" case refers to a planet with a semi-major axis half that of the current K2-18b, and ``nominal" refers to the nominal K2-18b. Both models assume T$_{\mathrm{int}}$ = 70 K. The P-T profiles for both cases are shown in Figure \ref{fig:TP}. Here we omit the 100-bar surface case results since they overlap with the no-surface results for both the hot and nominal cases. The dashed blue and red lines are respectively the equilibrium profiles for the cold and nominal cases. Other lines for the nominal case are colored in the same way as in Figure \ref{fig:HCN_NH3}. Lines for the hot case include: yellow line (1-bar surface), red line (10-bar surface), and thick red-gray line (no surface).}
\label{fig:hot_sens}
\end{figure}

We also test the sensitivity of our results to orbital distance, creating a model of a hotter sub-Neptune with K2-18b's physical parameters, except the orbital distance is half that of K2-18b (i.e., the incident stellar flux is increased by a factor of four), while keeping the same T$_{\mathrm{int}}$ (70 K) but increasing the strength of vertical mixing (K$_{zz}$) by a factor of 4 in keeping with the enhanced energy input. The full results for the hot case are shown in Figure \ref{fig:hot}. Comparison between the nominal and the hotter variant for selected species is shown in Figure \ref{fig:hot_sens}. Note that we did not test a colder variant because our current model does not include condensation, and water would condense over a large pressure range in the colder sub-Neptune atmosphere, leading to drastic changes in its abundance profiles and subsequent photochemical processes. 

The temperature difference between the hot case and the nominal case remains relatively constant in the upper atmosphere down to $\sim$1 bar, at which point the difference becomes smaller until the two P-T profiles coincide with each other at around $\sim$100 bar. Hotter temperatures lead to more effective thermochemical kinetics, allowing chemical reactions to better compete with vertical transport, in general, although vertical transport is also enhanced in the hot model. The net result is that in the no-surface case, thermochemical equilibrium persists to slightly lower pressures in the hot model than the nominal model. For example, the quench point for CH$_4$-CO-H$_2$O is at $\sim$10 bar for the hot case and $\sim$30 bar for the nominal case. At these respective quench points, the hot model has a smaller equilibrium mixing ratios of CH$_4$ and H$_2$O and a larger equilibrium mixing ratios of CO and CO$_2$ than the nominal model, and transport-induced quenching then preserves these differences as the quenched species are transported to the upper atmosphere. However, because the quench point for N$_2$-NH$_3$ is located deeper in the atmosphere (near $\sim$100 bar), where the thermal structure converges for the two models, the quench points and quenched mixing ratios of ammonia and N$_2$ are similar for the nominal and the hot cases, leading to similar N$_2$ and NH$_3$ VMR profiles when no surface is present. Note that we did not include the 100-bar surface case results in Figure \ref{fig:hot_sens}, as the abundance profiles for these models overlap with the no-surface case for all species for both the nominal and the hot models, because thermochemical kinetics is efficiently operating at 100-bar conditions to recycle the parent molecules and transport them back to the upper atmosphere. 

For the 10-bar surface case, the carbon and oxygen species in group 2 and 3 have similar abundance profiles as the no-surface case for both the hot and the nominal models, as the surface temperature is warm enough that thermochemical kinetic reactions help drive the C and O species toward their equilibrium abundances, such that the species that have been lost to photochemistry can be recycled (see section 3.2). The nitrogen species in group 1, in contrast, remain far from thermochemical equilibrium at the temperatures relevant to the 10-bar surface for both the hot and nominal models, so their mixing ratios are not tied to equilibrium at the surface. Instead, the VMR profiles of the nitrogen species are controlled largely by photochemical production/loss and transport. The warmer temperatures throughout the upper atmosphere for the hot case are more conducive to prompt recycling of NH$_3$ once it is photolyzed, through temperature-sensitive reactions such as NH$_2$ + H$_2$ $\rightarrow$ NH$_3$ + H, NH$_2$ + H$_2$O $\rightarrow$ NH$_3$ + OH, and NH$_2$ + CH$_4$ $\rightarrow$ NH$_3$ + CH$_3$. Moreover, other non-recycling pathways for NH$_2$ loss, such as NH$_2$ + CH$_3$ + M $\rightarrow$ CH$_3$NH$_2$ + M and NH$_2$ + NH$_2$ + M $\rightarrow$ N$_2$H$_4$ + M compete more effectively against the NH$_3$ recycling in the nominal model, leading to a lower ultimate steady-state VMR of NH$_3$ in the nominal model compared with the hot model.

The same scenario with respect to the nitrogen species occurs in the 1-bar surface cases, in terms of enhanced NH$_3$ in the hotter atmospheric model, as a result of more efficient ammonia recycling in the photochemically active region of the atmosphere when temperatures are higher. Thermochemical equilibrium is not recovered at the surface conditions relevant to the 1-bar surface for either the hot or nominal models, so the oxygen and carbon species in the 1-bar cases are also controlled by photochemical production, loss, and transport. H$_2$O behaves in a similar manner as NH$_3$ -- hotter atmospheric temperatures favor more efficient recycling of water once it is photolyzed, through the temperature-sensitive reactions such as OH + H$_2$ $\rightarrow$ H$_2$O + H -- so the hot 1-bar surface model has more H$_2$O than the nominal model. However, CH$_4$ is actually depleted and CO enhanced in the hot 1-bar surface case in comparison with the nominal case because the higher UV flux and higher H abundance help fuel the conversion of CH$_4$ into CO on the hot planet in comparison with the nominal planet. Despite the slightly enhanced CO abundance on the hot planet with the 1-bar surface, the high atmospheric temperatures significantly enhance the rate coefficient for the reaction H + CO$_2$ $\rightarrow$ CO + OH, which when combined with the larger H abundance due to the higher UV flux incident, cause a reduction in the abundance of CO$_2$ in the hot 1-bar surface case in comparison with the nominal 1-bar surface case.

\begin{figure}[h]
\centering
\includegraphics[width=\textwidth]{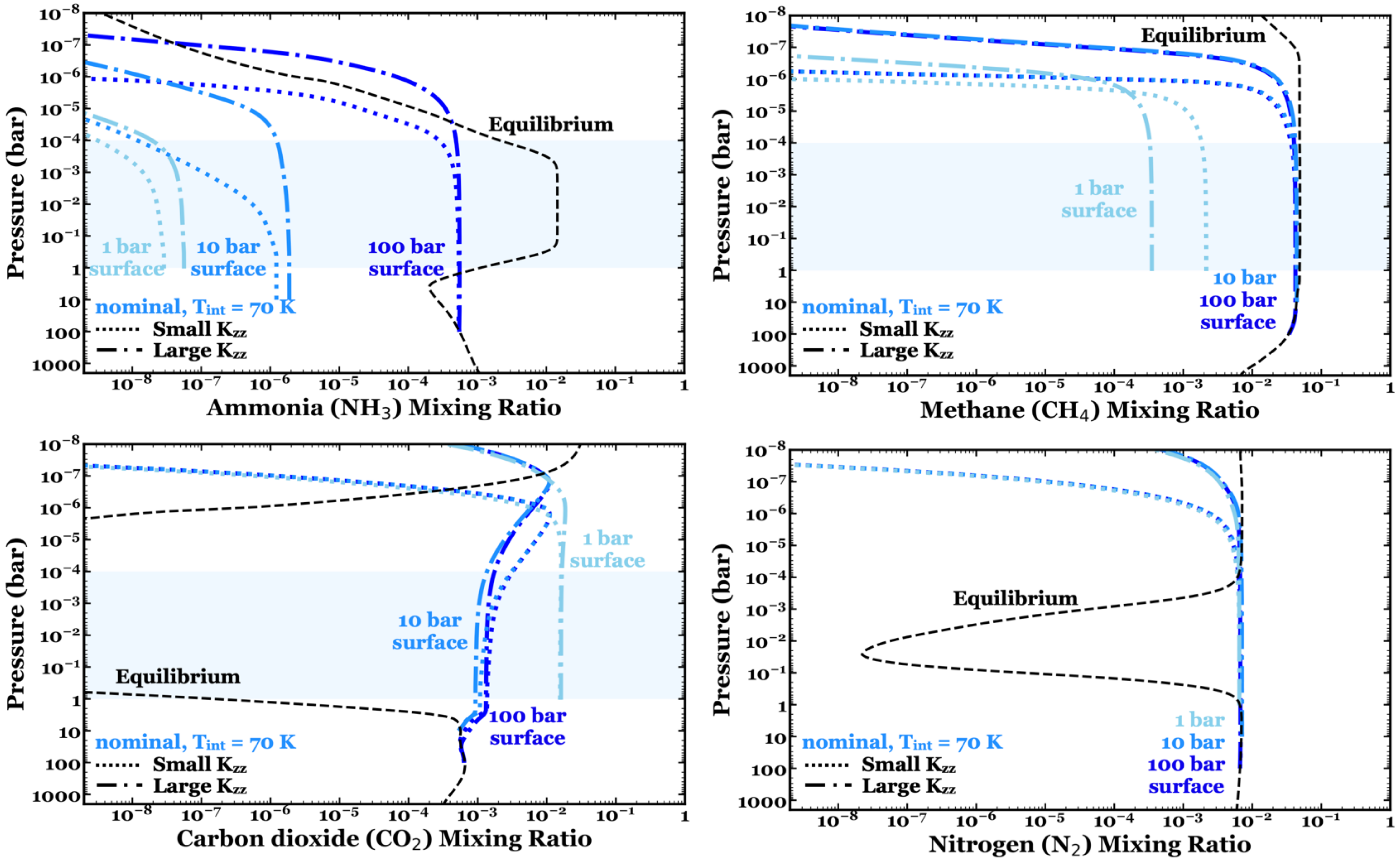}
\caption{The sensitivity of species abundances to different K$_{zz}$ profiles for the nominal case with T$_{\mathrm{int}}$ = 70 K. Here only the end member K$_{zz}$ profiles results are shown, the dotted lines are for the results of the small K$_{zz}$ profile (nominal profile, see Figure \ref{fig:TP}, divided by three) and the dash-dotted lines are for the large K$_{zz}$ profile (nominal profile multiplied by three). The lines are colored in the same way as in Figure \ref{fig:HCN_NH3}.}
\label{fig:kzz}
\end{figure}

Our model results are also sensitive to the assumed K$_{zz}$ profile, which can alter the pressure regions at which photochemistry, transport, and thermochemistry are most effective. Large K$_{zz}$ values lead to deeper quench points, which makes thermochemistry less effective at shallower pressures. Large K$_{zz}$ values also lead to more rigorous atmospheric mixing, allowing quenched species to be carried to higher altitudes leading to photochemistry being initiated at lower pressures. The pressure at which the main parent molecules are photolyzed can change their relative production rate (or recycling rate) and loss rate (conversion to new photochemical species). Large K$_{zz}$ values in the lower stratosphere can also cause depletion of photochemical products that were produced at higher altitudes, accelerating their transport and conversion back to the parent molecules.

Comparison between the model results using two end-member K$_{zz}$ profiles (so-called ``large" and ``small" K$_{zz}$ profiles) for selected species is shown in Figure \ref{fig:kzz}. In general, the main atmospheric species are not very sensitive to K$_{zz}$. The larger K$_{zz}$ causes a slightly deeper quench point for CO-CH$_4$-H$_2$O, which leads to a slight decrease in the quenched mixing ratios of CH$_4$ and H$_2$O and a larger increase in the quenched CO mixing ratio for the models whose surface is deep enough to capture this quench point (e.g., surface P $\geq$ 30 bar). The larger change in CO is caused by the larger vertical gradient of its equilibrium mixing ratio at depth in the atmosphere. A larger K$_{zz}$ also causes all the species to be carried to higher altitudes before molecular diffusion limits their abundance. This leads an increase in abundance for some photochemical products (e.g., C$_2$H$_2$, C4H$_2$, C$_6$H$_6$, O$_2$), but the larger K$_{zz}$ also results in to faster transport of photochemically produced species to the hot, deep lower atmosphere, where some of the more stable products are more easily destroyed, so the larger K$_{zz}$ actually leads to a decrease in the abundance of some species (e.g., C$_2$H$_6$, C$_3$H$_8$, HCN, CH$_3$CN, CH$_3$OH) at observable pressures. 

For the shallower surface cases (1-bar, 10-bar), variations in K$_{zz}$ have little effect on the overall stability and steady-state abundance profiles of the major parent molecules H$_2$O, CO, and CH$_4$ at observable pressure regions. The CO$_2$ abundance is slightly enhanced by the shift of the photolysis region to higher pressures that occurs with the lower K$_{zz}$ case. The higher-pressure photolysis region in the low K$_{zz}$ case also enhances the conversion of NH$_3$ into HCN, so NH$_3$ is notably depleted and HCN is correspondingly enhanced at observable pressure levels in the low K$_{zz}$ case. Although CH$_4$ also participates in this HCN formation, methane is much more abundant than ammonia to begin with (because of the recycling of hydrocarbons to methane at surface conditions, see section 3.2), so the loss to HCN has little effect on the CH$_4$ abundance. Variations in K$_{zz}$ do not affect our main conclusion that species in group 1 such as NH$_3$ and HCN are still sensitive to 1-bar and 10-bar surfaces and have different mixing ratios compared to the no-surface case in the observable part of the atmosphere. Species in group 2 and 3 are still only sensitive to 1-bar surfaces.

Overall, our main qualitative conclusions in Section 3.2 remain true for species in different groups with the changed T$_{\mathrm{int}}$, semimajor axis, and K$_{zz}$ profiles. A very shallow, cool 1-bar surface would generally lead to decreased levels of NH$_3$, HCN, CH$_4$, C$_x$H$_y$, H$_2$O, and increased levels of CO compared to the no-surface case. However, the increased levels of CO$_2$ for the 1-bar surface case become more muted for the hotter planet at half K2-18b's orbital distance. A 10-bar surface would lead to decreased levels of NH$_3$ and HCN, but deeper surfaces at pressures beyond 100-bar would lead to the same results as the no-surface case.


\begin{thebibliography}{}
\bibitem[Ag{\'u}ndez et al.(2014)]{2014ApJ...781...68A} Ag{\'u}ndez, M., Venot, O., Selsis, F., et al.\ 2014, \apj, 781, 68. doi:10.1088/0004-637X/781/2/68
\bibitem[Allen et al.(1980)]{1980ApJ...242L.125A} Allen, M., Yung, Y.~L., \& Pinto, J.~P.\ 1980, \apjl, 242, L125. doi:10.1086/183416
\bibitem[Anderson et al.(2018)]{2018SSRv..214..125A} Anderson, C.~M., Samuelson, R.~E., \& Nna-Mvondo, D.\ 2018, \ssr, 214, 125. doi:10.1007/s11214-018-0559-5
\bibitem[Atreya et al.(2006)]{2006P&SS...54.1177A} Atreya, S.~K., Adams, E.~Y., Niemann, H.~B., et al.\ 2006, \planss, 54, 1177. doi:10.1016/j.pss.2006.05.028
\bibitem[Atreya et al.(2010)]{2010tfch.book..177A} Atreya, S.~K., Lorenz, R.~D., \& Waite, J.~H.\ 2010, Titan from Cassini-Huygens, 177. doi:10.1007/978-1-4020-9215-2$\_$7
\bibitem[Batalha(2014)]{2014PNAS..11112647B} Batalha, N.~M.\ 2014, Proceedings of the National Academy of Science, 111, 12647. doi:10.1073/pnas.1304196111
\bibitem[Bean et al.(2021)]{2021JGRE..12606639B} Bean, J.~L., Raymond, S.~N., \& Owen, J.~E.\ 2021, Journal of Geophysical Research (Planets), 126, e06639. doi:10.1029/2020JE006639
\bibitem[Benneke et al.(2017)]{2017ApJ...834..187B} Benneke, B., Werner, M., Petigura, E., et al.\ 2017, \apj, 834, 187. doi:10.3847/1538-4357/834/2/187
\bibitem[Benneke et al.(2019a)]{2019ApJ...887L..14B} Benneke, B., Wong, I., Piaulet, C., et al.\ 2019, \apjl, 887, L14. doi:10.3847/2041-8213/ab59dc
\bibitem[Benneke et al.(2019b)]{2019NatAs...3..813B} Benneke, B., Knutson, H.~A., Lothringer, J., et al.\ 2019, Nature Astronomy, 3, 813. doi:10.1038/s41550-019-0800-5
\bibitem[B{\'e}zard et al.(2020)]{2020arXiv201110424B} B{\'e}zard, B., Charnay, B., \& Blain, D.\ 2020, arXiv:2011.10424
\bibitem[Blain et al.(2021)]{2021A&A...646A..15B} Blain, D., Charnay, B., \& B{\'e}zard, B.\ 2021, \aap, 646, A15. doi:10.1051/0004-6361/202039072
\bibitem[Burrows \& Sharp(1999)]{1999ApJ...512..843B} Burrows, A. \& Sharp, C.~M.\ 1999, \apj, 512, 843. doi:10.1086/306811
\bibitem[Chachan \& Stevenson(2018)]{2018ApJ...854...21C} Chachan, Y. \& Stevenson, D.~J.\ 2018, \apj, 854, 21. doi:10.3847/1538-4357/aaa459
\bibitem[Chachan et al.(2020)]{2020AJ....160..201C} Chachan, Y., Jontof-Hutter, D., Knutson, H.~A., et al.\ 2020, \aj, 160, 201. doi:10.3847/1538-3881/abb23a
\bibitem[Changeat et al.(2020)]{2020arXiv200301486C} Changeat, Q., Edwards, B., Al-Refaie, A.~F., et al.\ 2020, arXiv:2003.01486
\bibitem[Charnay et al.(2021)]{2021a&a201111553C} Charnay, B., Blain, D., B{\'e}zard, B., et al.\ 2020, \aap, 646, A171. doi:10.1051/0004-6361/202039525
\bibitem[Cloutier et al.(2017)]{2017A&A...608A..35C} Cloutier, R., Astudillo-Defru, N., Doyon, R., et al.\ 2017, \aap, 608, A35. doi:10.1051/0004-6361/201731558
\bibitem[Cloutier et al.(2019)]{2019A&A...621A..49C} Cloutier, R., Astudillo-Defru, N., Doyon, R., et al.\ 2019, \aap, 621, A49. doi:10.1051/0004-6361/201833995
\bibitem[Coates et al.(2007)]{2007GeoRL..3422103C} Coates, A.~J., Crary, F.~J., Lewis, G.~R., et al.\ 2007, \grl, 34, L22103. doi:10.1029/2007GL030978
\bibitem[dos Santos et al.(2020)]{2020A&A...634L...4D} dos Santos, L.~A., Ehrenreich, D., Bourrier, V., et al.\ 2020, \aap, 634, L4. doi:10.1051/0004-6361/201937327
\bibitem[Elkins-Tanton \& Seager(2008)]{2008ApJ...685.1237E} Elkins-Tanton, L.~T. \& Seager, S.\ 2008, \apj, 685, 1237. doi:10.1086/591433
\bibitem[Fressin et al.(2013)]{2013ApJ...766...81F} Fressin, F., Torres, G., Charbonneau, D., et al.\ 2013, \apj, 766, 81. doi:10.1088/0004-637X/766/2/81
\bibitem[Fortney et al.(2005)]{2005ApJ...627L..69F} Fortney, J.~J., Marley, M.~S., Lodders, K., et al.\ 2005, \apjl, 627, L69. doi:10.1086/431952
\bibitem[Fortney et al.(2008)]{2008ApJ...678.1419F} Fortney, J.~J., Lodders, K., Marley, M.~S., et al.\ 2008, \apj, 678, 1419. doi:10.1086/528370
\bibitem[Fortney et al.(2013)]{2013ApJ...775...80F} Fortney, J.~J., Mordasini, C., Nettelmann, N., et al.\ 2013, \apj, 775, 80. doi:10.1088/0004-637X/775/1/80
\bibitem[Fortney et al.(2020)]{2020AJ....160..288F} Fortney, J.~J., Visscher, C., Marley, M.~S., et al.\ 2020, \aj, 160, 288. doi:10.3847/1538-3881/abc5bd
\bibitem[France et al.(2016)]{2016ApJ...820...89F} France, K., Loyd, R.~O.~P., Youngblood, A., et al.\ 2016, \apj, 820, 89. doi:10.3847/0004-637X/820/2/89
\bibitem[Fulton \& Petigura(2018)]{2018AJ....156..264F} Fulton, B.~J. \& Petigura, E.~A.\ 2018, \aj, 156, 264. doi:10.3847/1538-3881/aae828
\bibitem[Fulton et al.(2017)]{2017AJ....154..109F} Fulton, B.~J., Petigura, E.~A., Howard, A.~W., et al.\ 2017, \aj, 154, 109. doi:10.3847/1538-3881/aa80eb
\bibitem[Gaillard \& Scaillet(2014)]{2014E&PSL.403..307G} Gaillard, F. \& Scaillet, B.\ 2014, Earth and Planetary Science Letters, 403, 307. doi:10.1016/j.epsl.2014.07.009
\bibitem[Ginzburg et al.(2018)]{2018MNRAS.476..759G} Ginzburg, S., Schlichting, H.~E., \& Sari, R.\ 2018, \mnras, 476, 759. doi:10.1093/mnras/sty290
\bibitem[Glein(2015)]{2015Icar..250..570G} Glein, C.~R.\ 2015, \icarus, 250, 570. doi:10.1016/j.icarus.2015.01.001
\bibitem[Gordon \& McBride(1984)]{1984cpcc.book.....G} Gordon, S., \& McBride, B.\ 1984, NASA reference publication, 1311.
\bibitem[Guo et al.(2020)]{2020AJ....159..239G} Guo, X., Crossfield, I.~J.~M., Dragomir, D., et al.\ 2020, \aj, 159, 239. doi:10.3847/1538-3881/ab8815
\bibitem[Hayes et al.(2018)]{2018NatGe..11..306H} Hayes, A.~G., Lorenz, R.~D., \& Lunine, J.~I.\ 2018, Nature Geoscience, 11, 306. doi:10.1038/s41561-018-0103-y
\bibitem[He et al.(2018a)]{2018AJ....156...38H} He, C., H{\"o}rst, S.~M., Lewis, N.~K., et al.\ 2018, \aj, 156, 38. doi:10.3847/1538-3881/aac883
\bibitem[He et al.(2018b)]{2018ApJ...856L...3H} He, C., H{\"o}rst, S.~M., Lewis, N.~K., et al.\ 2018, \apjl, 856, L3. doi:10.3847/2041-8213/aab42b
\bibitem[He et al.(2019)]{2019ECS.....3...39H} He, C., H{\"o}rst, S.~M., Lewis, N.~K., et al.\ 2019, ACS Earth and Space Chemistry, 3, 39. doi:10.1021/acsearthspacechem.8b00133
\bibitem[He et al.(2020a)]{2020NatAs...4..986H} He, C., H{\"o}rst, S.~M., Lewis, N.~K., et al.\ 2020, Nature Astronomy, 4, 986. doi:10.1038/s41550-020-1072-9
\bibitem[He et al.(2020b)]{2020PSJ.....1...51H} He, C., H{\"o}rst, S.~M., Lewis, N.~K., et al.\ 2020, The Planetary Science Journal, 1, 51. doi:10.3847/PSJ/abb1a4
\bibitem[H{\"o}rst et al.(2018)]{2018NatAs...2..303H} H{\"o}rst, S.~M., He, C., Lewis, N.~K., et al.\ 2018, Nature Astronomy, 2, 303. doi:10.1038/s41550-018-0397-0
\bibitem[Irwin(2009)]{2009book....156..264F} Irwin, P.\ 2009, Giant planets of our solar system: atmospheres, composition, and structure (2nd ed,; Chichester, UK, Praxis Publishing Ltd.). 
\bibitem[Kite et al.(2019)]{2019ApJ...887L..33K} Kite, E.~S., Fegley, B., Schaefer, L., et al.\ 2019, \apjl, 887, L33. doi:10.3847/2041-8213/ab59d9
\bibitem[Kite et al.(2020)]{2020ApJ...891..111K} Kite, E.~S., Fegley, B., Schaefer, L., et al.\ 2020, \apj, 891, 111. doi:10.3847/1538-4357/ab6ffb
\bibitem[Knutson et al.(2011)]{2011ApJ...735...27K} Knutson, H.~A., Madhusudhan, N., Cowan, N.~B., et al.\ 2011, \apj, 735, 27. doi:10.1088/0004-637X/735/1/27
\bibitem[Knutson et al.(2014a)]{2014Natur.505...66K} Knutson, H.~A., Benneke, B., Deming, D., et al.\ 2014, \nat, 505, 66. doi:10.1038/nature12887
\bibitem[Knutson et al.(2014b)]{2014ApJ...794..155K} Knutson, H.~A., Dragomir, D., Kreidberg, L., et al.\ 2014, \apj, 794, 155. doi:10.1088/0004-637X/794/2/155
\bibitem[Koll et al.(2019)]{2019ApJ...886..140K} Koll, D.~D.~B., Malik, M., Mansfield, M., et al.\ 2019, \apj, 886, 140. doi:10.3847/1538-4357/ab4c91
\bibitem[Kreidberg et al.(2014)]{2014Natur.505...69K} Kreidberg, L., Bean, J.~L., D{\'e}sert, J.-M., et al.\ 2014, \nat, 505, 69. doi:10.1038/nature12888
\bibitem[Kreidberg et al.(2018)]{2018ApJ...858L...6K} Kreidberg, L., Line, M.~R., Thorngren, D., et al.\ 2018, \apjl, 858, L6. doi:10.3847/2041-8213/aabfce
\bibitem[Kreidberg et al.(2019)]{2019Natur.573...87K} Kreidberg, L., Koll, D.~D.~B., Morley, C., et al.\ 2019, \nat, 573, 87. doi:10.1038/s41586-019-1497-4
\bibitem[Lanotte et al.(2014)]{2014A&A...572A..73L} Lanotte, A.~A., Gillon, M., Demory, B.-O., et al.\ 2014, \aap, 572, A73. doi:10.1051/0004-6361/201424373
\bibitem[Libby-Roberts et al.(2020)]{2020AJ....159...57L} Libby-Roberts, J.~E., Berta-Thompson, Z.~K., D{\'e}sert, J.-M., et al.\ 2020, \aj, 159, 57. doi:10.3847/1538-3881/ab5d36
\bibitem[Line et al.(2011)]{2011ApJ...738...32L} Line, M.~R., Vasisht, G., Chen, P., et al.\ 2011, \apj, 738, 32. doi:10.1088/0004-637X/738/1/32
\bibitem[Lodders \& Fegley(2002)]{2002Icar..155..393L} Lodders, K. \& Fegley, B.\ 2002, \icarus, 155, 393. doi:10.1006/icar.2001.6740
\bibitem[Lodders(2010)]{2010ASSP...16..379L} Lodders, K.\ 2010, Astrophysics and Space Science Proceedings, 16, 379. doi:10.1007/978-3-642-10352-0$\_$8
\bibitem[Lopez \& Fortney(2013)]{2013ApJ...776....2L} Lopez, E.~D. \& Fortney, J.~J.\ 2013, \apj, 776, 2. doi:10.1088/0004-637X/776/1/2
\bibitem[Lunine \& Stevenson(1987)]{1987Icar...70...61L} Lunine, J.~I. \& Stevenson, D.~J.\ 1987, \icarus, 70, 61. doi:10.1016/0019-1035(87)90075-3
\bibitem[Madhusudhan \& Seager(2011)]{2011ApJ...729...41M} Madhusudhan, N. \& Seager, S.\ 2011, \apj, 729, 41. doi:10.1088/0004-637X/729/1/41
\bibitem[Madhusudhan et al.(2020)]{2020ApJ...891L...7M} Madhusudhan, N., Nixon, M.~C., Welbanks, L., et al.\ 2020, \apjl, 891, L7. doi:10.3847/2041-8213/ab7229
\bibitem[Malik et al.(2019)]{2019ApJ...886..142M} Malik, M., Kempton, E.~M.-R., Koll, D.~D.~B., et al.\ 2019, \apj, 886, 142. doi:10.3847/1538-4357/ab4a05
\bibitem[Malsky \& Rogers(2020)]{2020ApJ...896...48M} Malsky, I. \& Rogers, L.~A.\ 2020, \apj, 896, 48. doi:10.3847/1538-4357/ab873f
\bibitem[Mandt et al.(2014)]{2014ApJ...788L..24M} Mandt, K.~E., Mousis, O., Lunine, J., et al.\ 2014, \apjl, 788, L24. doi:10.1088/2041-8205/788/2/L24
\bibitem[Mansfield et al.(2019)]{2019ApJ...886..141M} Mansfield, M., Kite, E.~S., Hu, R., et al.\ 2019, \apj, 886, 141. doi:10.3847/1538-4357/ab4c90
\bibitem[Marley \& McKay(1999)]{1999Icar..138..268M} Marley, M.~S. \& McKay, C.~P.\ 1999, \icarus, 138, 268. doi:10.1006/icar.1998.6071
\bibitem[May \& Rauscher(2020)]{2020ApJ...893..161M} May, E.~M. \& Rauscher, E.\ 2020, \apj, 893, 161. doi:10.3847/1538-4357/ab838b
\bibitem[McKay et al.(1989)]{1989Icar...80...23M} McKay, C.~P., Pollack, J.~B., \& Courtin, R.\ 1989, \icarus, 80, 23. doi:10.1016/0019-1035(89)90160-7
\bibitem[Moran et al.(2020)]{2020PSJ.....1...17M} Moran, S.~E., H{\"o}rst, S.~M., Vuitton, V., et al.\ 2020, The Planetary Science Journal, 1, 17. doi:10.3847/PSJ/ab8eae
\bibitem[Mordasini(2020)]{2020A&A...638A..52M} Mordasini, C.\ 2020, \aap, 638, A52. doi:10.1051/0004-6361/201935541
\bibitem[Morley et al.(2012)]{2012ApJ...756..172M} Morley, C.~V., Fortney, J.~J., Marley, M.~S., et al.\ 2012, \apj, 756, 172. doi:10.1088/0004-637X/756/2/172
\bibitem[Morley et al.(2017)]{2017AJ....153...86M} Morley, C.~V., Knutson, H., Line, M., et al.\ 2017, \aj, 153, 86. doi:10.3847/1538-3881/153/2/86
\bibitem[Moses(2014)]{2014RSPTA.37230073M} Moses, J.~I.\ 2014, Philosophical Transactions of the Royal Society of London Series A, 372, 20130073. doi:10.1098/rsta.2013.0073
\bibitem[Moses et al.(2005)]{2005JGRE..110.8001M} Moses, J.~I., Fouchet, T., B{\'e}zard, B., et al.\ 2005, Journal of Geophysical Research (Planets), 110, E08001. doi:10.1029/2005JE002411
\bibitem[Moses et al.(2010)]{2010FaDi..147..103M} Moses, J.~I., Visscher, C., Keane, T.~C., et al.\ 2010, Faraday Discussions, 147, 103. doi:10.1039/c003954c
\bibitem[Moses et al.(2011)]{2011ApJ...737...15M} Moses, J.~I., Visscher, C., Fortney, J.~J., et al.\ 2011, \apj, 737, 15. doi:10.1088/0004-637X/737/1/15
\bibitem[Moses et al.(2013)]{2013ApJ...777...34M} Moses, J.~I., Line, M.~R., Visscher, C., et al.\ 2013, \apj, 777, 34. doi:10.1088/0004-637X/777/1/34
\bibitem[Moses et al.(2016)]{2016ApJ...829...66M} Moses, J.~I., Marley, M.~S., Zahnle, K., et al.\ 2016, \apj, 829, 66. doi:10.3847/0004-637X/829/2/66
\bibitem[Moses et al.(2020)]{2020ptrsa200611367M} Moses, J.~I., Cavalie, T., Fletcher, L.~N., et al.\ 2020, Philosophical Transactions of the Royal Society A, 378, 2187. doi: 10.1098/rsta.2019.0477.
\bibitem[Mousis et al.(2020)]{2020ApJ...896L..22M} Mousis, O., Deleuil, M., Aguichine, A., et al.\ 2020, \apjl, 896, L22. doi:10.3847/2041-8213/ab9530
\bibitem[Niemann et al.(2005)]{2005Natur.438..779N} Niemann, H.~B., Atreya, S.~K., Bauer, S.~J., et al.\ 2005, \nat, 438, 779. doi:10.1038/nature04122
\bibitem[Niemann et al.(2010)]{2010JGRE..11512006N} Niemann, H.~B., Atreya, S.~K., Demick, J.~E., et al.\ 2010, Journal of Geophysical Research (Planets), 115, E12006. doi:10.1029/2010JE003659
\bibitem[Nixon et al.(2010)]{2010FaDi..147...65N} Nixon, C.~A., Achterberg, R.~K., Teanby, N.~A., et al.\ 2010, Faraday Discussions, 147, 65. doi:10.1039/c003771k
\bibitem[Owen \& Wu(2013)]{2013ApJ...775..105O} Owen, J.~E. \& Wu, Y.\ 2013, \apj, 775, 105. doi:10.1088/0004-637X/775/2/105
\bibitem[Pearl \& Conrath(1991)]{1991JGR....9618921P} Pearl, J.~C. \& Conrath, B.~J.\ 1991, \jgr, 96, 18921. doi:10.1029/91JA01087
\bibitem[Piette \& Madhusudhan(2020)]{2020ApJ...904..154P} Piette, A.~A.~A. \& Madhusudhan, N.\ 2020, \apj, 904, 154. doi:10.3847/1538-4357/abbfb1
\bibitem[Prinn \& Barshay(1977)]{1977Sci...198.1031P} Prinn, R.~G. \& Barshay, S.~S.\ 1977, Science, 198, 1031. doi:10.1126/science.198.4321.1031
\bibitem[Schaefer et al.(2016)]{2016ApJ...829...63S} Schaefer, L., Wordsworth, R.~D., Berta-Thompson, Z., et al.\ 2016, \apj, 829, 63. doi:10.3847/0004-637X/829/2/63
\bibitem[Scheucher et al.(2020)]{2020ApJ...898...44S} Scheucher, M., Wunderlich, F., Grenfell, J.~L., et al.\ 2020, \apj, 898, 44. doi:10.3847/1538-4357/ab9084
\bibitem[Sharp \& Burrows(2007)]{2007ApJS..168..140S} Sharp, C.~M. \& Burrows, A.\ 2007, \apjs, 168, 140. doi:10.1086/508708
\bibitem[Stevenson et al.(2010)]{2010Natur.464.1161S} Stevenson, K.~B., Harrington, J., Nymeyer, S., et al.\ 2010, \nat, 464, 1161. doi:10.1038/nature09013
\bibitem[Stevenson et al.(2012)]{2012ApJ...755....9S} Stevenson, K.~B., Harrington, J., Lust, N.~B., et al.\ 2012, \apj, 755, 9. doi:10.1088/0004-637X/755/1/9
\bibitem[Stone(1976)]{1976jupiter...26..906S} Stone, P.~H.\ 1976, in Jupiter, ed. T. Gehrels (Tucson, AZ: Univ. Arizona Press), 586.
\bibitem[Strobel(1969)]{1969JAtS...26..906S} Strobel, D.~F.\ 1969, Journal of Atmospheric Sciences, 26, 906. doi:10.1175/1520-0469(1969)026<0906:TPOMIT>2.0.CO;2
\bibitem[Strobel(1973)]{1973JAtS...30.1205S} Strobel, D.~F.\ 1973, Journal of Atmospheric Sciences, 30, 1205. doi:10.1175/1520-0469(1973)030<1205:TPONIT>2.0.CO;2
\bibitem[Teanby et al.(2013)]{2013P&SS...75..136T} Teanby, N.~A., Irwin, P.~G.~J., Nixon, C.~A., et al.\ 2013, \planss, 75, 136. doi:10.1016/j.pss.2012.11.008
\bibitem[Thompson et al.(2021)]{2021natas...sdsY} Thompson, M.~A., Telus, M., Schaefer, L., et al.\ 2021, in revision.
\bibitem[Tobie et al.(2006)]{2006Natur.440...61T} Tobie, G., Lunine, J.~I., \& Sotin, C.\ 2006, \nat, 440, 61. doi:10.1038/nature04497
\bibitem[Toon et al.(1989)]{1989JGR....9416287T} Toon, O.~B., McKay, C.~P., Ackerman, T.~P., et al.\ 1989, \jgr, 94, 16287. doi:10.1029/JD094iD13p16287
\bibitem[Tsiaras et al.(2019)]{2019NatAs...3.1086T} Tsiaras, A., Waldmann, I.~P., Tinetti, G., et al.\ 2019, Nature Astronomy, 3, 1086. doi:10.1038/s41550-019-0878-9
\bibitem[Van Eylen et al.(2018)]{2018MNRAS.479.4786V} Van Eylen, V., Agentoft, C., Lundkvist, M.~S., et al.\ 2018, \mnras, 479, 4786. doi:10.1093/mnras/sty1783
\bibitem[Venot et al.(2012)]{2012A&A...546A..43V} Venot, O., H{\'e}brard, E., Ag{\'u}ndez, M., et al.\ 2012, \aap, 546, A43. doi:10.1051/0004-6361/201219310
\bibitem[Visscher et al.(2010)]{2010ApJ...716.1060V} Visscher, C., Lodders, K., \& Fegley, B.\ 2010, \apj, 716, 1060. doi:10.1088/0004-637X/716/2/1060
\bibitem[Visscher \& Moses(2011)]{2011ApJ...738...72V} Visscher, C. \& Moses, J.~I.\ 2011, \apj, 738, 72. doi:10.1088/0004-637X/738/1/72
\bibitem[Vuitton et al.(2007)]{2007Icar..191..722V} Vuitton, V., Yelle, R.~V., \& McEwan, M.~J.\ 2007, \icarus, 191, 722. doi:10.1016/j.icarus.2007.06.023
\bibitem[Vuitton et al.(2009)]{2009P&SS...57.1558V} Vuitton, V., Lavvas, P., Yelle, R.~V., et al.\ 2009, \planss, 57, 1558. doi:10.1016/j.pss.2009.04.004
\bibitem[Vuitton et al.(2021)]{2021PSJ.....2....2V} Vuitton, V., Moran, S.~E., He, C., et al.\ 2021, The Planetary Science Journal, 2, 2. doi:10.3847/PSJ/abc558
\bibitem[Wang et al.(2017)]{2017ApJ...850..199W} Wang, D., Miguel, Y., \& Lunine, J.\ 2017, \apj, 850, 199. doi:10.3847/1538-4357/aa978e
\bibitem[Wogan et al.(2020)]{2020PSJ.....1...58W} Wogan, N., Krissansen-Totton, J., \& Catling, D.~C.\ 2020, The Planetary Science Journal, 1, 58. doi:10.3847/PSJ/abb99e
\bibitem[Yelle et al.(2010)]{2010FaDi..147...31Y} Yelle, R.~V., Vuitton, V., Lavvas, P., et al.\ 2010, Faraday Discussions, 147, 31. doi:10.1039/c004787m
\bibitem[Yung \& Demore(1999)]{1999ppa..conf.....Y} Yung, Y.~L. \& Demore, W.~B.\ 1999, Photochemistry of planetary atmospheres / Yuk L. Yung
\bibitem[Yung et al.(1984)]{1984ApJS...55..465Y} Yung, Y.~L., Allen, M., \& Pinto, J.~P.\ 1984, \apjs, 55, 465. doi:10.1086/190963
\bibitem[Yu et al.(2021)]{2021natas...55..465Y} Yu, X., He, C., Zhang, X., et al.\ 2021, in revision.
\bibitem[Zahnle et al.(2009)]{2009ApJ...701L..20Z} Zahnle, K., Marley, M.~S., Freedman, R.~S., et al.\ 2009, \apjl, 701, L20. doi:10.1088/0004-637X/701/1/L20
\bibitem[Zeng et al.(2019)]{2019PNAS..116.9723Z} Zeng, L., Jacobsen, S.~B., Sasselov, D.~D., et al.\ 2019, Proceedings of the National Academy of Science, 116, 9723. doi:10.1073/pnas.1812905116
\end{thebibliography}
\end{document}